\documentclass[12pt]{article}
\usepackage{amsthm}
\usepackage{amsmath}
\usepackage{natbib}
\usepackage{graphicx}
\usepackage{geometry}
\usepackage{mathrsfs}  
\usepackage{lipsum}  
\usepackage{booktabs}
\usepackage{float}
\usepackage{amsfonts,amssymb}
\usepackage{color, xcolor}
\usepackage{scalerel}
\usepackage{multirow}
\usepackage{subcaption}
\usepackage{mathtools}
\usepackage{authblk}
\usepackage{tikz}
\usetikzlibrary{positioning, arrows.meta}
\usepackage{xr}
\usepackage[nodisplayskipstretch]{setspace}
\def\spacingset#1{\renewcommand{\baselinestretch}%
		{#1}\small\normalsize} \spacingset{1}

\usepackage{algorithm,algorithmicx,algpseudocode}
\graphicspath{{./fig/}}

\usepackage{geometry}
\newtheorem{theorem}{\bf{Theorem}}
\newtheorem{corollary}{\bf{Corollary}}
\newtheorem{remark}{\bf{Remark}}

\def\T{{\mathrm{\scriptscriptstyle T} }}
\def\diag{{\mathrm{diag} }}
\def\cov{{\mathrm{cov}}}
\def\var{\mathrm{var}}

\def\hE{\hat{E}}
\def\hcov{\widehat {{\mathrm{cov}}}}

\newcommand{\beps}{\boldsymbol{\epsilon}}
\newcommand{\bX}{\boldsymbol{X}}
\newcommand{\bY}{\boldsymbol{Y}}

\newcommand{\bG}{\boldsymbol{G}}

\newcommand{\fb}{\boldsymbol{f}}
\newcommand{\bw}{\boldsymbol{w}}
\newcommand{\bxi}{\boldsymbol{\xi}}

\newcommand{\bgamma}{\boldsymbol{\gamma}}

\newcommand{\bbF}{\mathbf{F}}

\newcommand{\bbI}{\mathbf{I}}
\newcommand{\bbM}{\mathbf{M}}
\newcommand{\bbH}{\mathbf{H}}

\newcommand{\bbA}{\mathbf{A}}
\newcommand{\bbPi}{\mathbf{\Pi}}
\newcommand{\bbXi}{\mathbf{\Xi}}

\newcommand{\bbSig}{\mathbf{\Sigma}}

\newcommand{\citeg}[1]{\citep[e.g.,][]{#1}}

\newcommand{\bzero}{\boldsymbol{0}}

\newcommand{\bone}{\boldsymbol{1}}

\newcommand{\btheta}{\boldsymbol{\theta}}

\newcommand{\tf}{\boldsymbol{\tilde{f}}}

\newcommand{\hw}{\hat{\boldsymbol{w}}}

\newcommand{\hbeta}{\hat{\beta}}

\newcommand{\htheta}{\hat{\boldsymbol{\theta}}}

\newcommand{\tbeta}{\tilde{\beta}}

\newcommand{\sD}{\mathscr{D}}

\newcommand{\cI}{\mathcal{I}}
\newcommand{\cT}{\scaleto{\mathcal{T}\mathstrut}{5pt}}
\newcommand{\cS}{\scaleto{\mathcal{S}\mathstrut}{5pt}}
\newcommand{\KNIVES}{\scaleto{{\rm KNIVES}\mathstrut}{5pt}}

\newcommand{\FT}{\scaleto{\rm FT\mathstrut}{5pt}}

\newcommand\blfootnote[1]{%
	\begingroup
	\renewcommand\thefootnote{}\footnote{#1}%
	\addtocounter{footnote}{-1}%
	\endgroup
}

\usepackage{url}

\title{
Instrumental Variable Analysis in Underrepresented Subpopulations Powered by Knowledge Transfer
}
\author[a]{Ruoyu Wang}

\author[b]{Zijian Xu}
\author[c$*$]{Zijian Guo}
\author[d,e$*$]{Xihong Lin}
\author[f$*$]{Molei Liu}

\affil[a]{Department of Statistics and Data Science, Tsinghua University}
\affil[b]{Department of Health Outcomes \& Biomedical Informatics, University of Florida}
\affil[c]{Center for Data Science, Zhejiang University}
\affil[d]{Department of Biostatistics, Harvard T.H. Chan School of Public Health}
\affil[e]{Department of Statistics, Harvard University}
\affil[f]{Department of Biostatistics, Peking University}
\date{}
\begin{document}
\pagenumbering{gobble}
\maketitle
\blfootnote{$*$ Co-corresponding authors. Emails: zijguo@zju.edu.cn; xlin@hsph.harvard.edu, moleiliu@bjmu.edu.cn} 
\begin{abstract}
\noindent Instrumental variable (IV) analysis in underrepresented target populations suffers from low statistical efficiency and weak IV bias due to limited sample size. Leveraging external knowledge from a source population with larger samples offers an appealing solution to this problem. We propose in this paper the {\bf KN}owledge-powered {\bf IV} {\bf E}stimator for underrepresented {\bf S}ubpopulations (KNIVES), a framework that enhances the IV analysis for underrepresented subpopulations by transferring knowledge from external source population data. KNIVES constructs a knowledge-transferred IV score by combining multiple candidate models through an effective signal-to-noise ratio criterion tailored to the downstream IV estimator, rather than prediction accuracy alone. To address weak associations between the IV score and the exposure, as well as a high correlation among candidate models in the target population, KNIVES employs a bias correction technique to mitigate the resulting bias and ensure robust performance. Theoretical investigation demonstrates that our method is robust to weak IV bias and achieves improved efficiency compared to the causal effect estimator based on any single candidate model. In addition, KNIVES avoids negative transfer by guaranteeing that its asymptotic variance is not larger than that of the estimator relying solely on target population data. Extensive numerical studies demonstrate the superior finite-sample performance of our method over existing methods. Two Mendelian randomization studies on ethnic minority subgroups in UK Biobank data further illustrate the advantages of our method. 
\end{abstract}
\noindent%
\vfill

\newpage
\spacingset{1.8}
\pagenumbering{arabic}

\section{Introduction}
\subsection{Motivation}




Causal inference is of substantial practical interest but remains challenging in observational studies due to unmeasured confounding. The instrumental variable (IV) method is one of the most widely used approaches for causal inference in the presence of unmeasured confounding. Mendelian randomization (MR), a prominent example of the IV method, uses genetic variants as instruments and has become a popular and powerful tool in epidemiological studies \citep{burgess2019guidelines}. 
However, in many applications, the causal effect of exposure may vary between subpopulations \citep{zheng2022multi}. {Estimating such subpopulation-specific causal effects is challenged by limited sample sizes of underrepresented demographic groups, weak instrument effects, and complex IV--exposure relationships. These problems can substantially reduce the estimation efficiency and lead to weak IV bias if not handled carefully \citep{newey1993eff,bound1995problems}. Our primary goal in this paper is to develop a transfer IV method that improves  IV analysis in the underrepresented target population by transferring useful IV--exposure information from large source populations.}

The UK Biobank provides a concrete example. Although the cohort contains approximately 500,000 participants, with more than 420,000 European-ancestry (EUR) individuals, the available sample sizes for non-European subpopulations are much smaller: about 7,000 for African ancestry (AFR), similarly limited for South Asian (SAS), and even smaller for some admixed ancestry groups. 
{Such limited sample sizes can result in IV models that explain only a small proportion of the exposure variation.} For example,  
our preliminary analysis on UK Biobank reveals very weak IV strength for body mass index (BMI) in both the AFR and SAS subgroups; see Table \ref{table: BMI-SBP}. This weak-IV issue leads to reduced power, unstable estimation, and bias. To address this challenge, we propose a transfer IV analysis framework that leverages the IV--exposure information from a much larger source population, e.g., the EUR data from the UK Biobank or external large-scale studies, to improve IV analysis in the underrepresented target population.

These larger source data sets, together with pretrained models from existing resources such as the polygenic score (PGS) catalog \citep{lambert2021polygenic}, provide rich external knowledge to strengthen IV--exposure modeling in studies of underrepresented subpopulations. {The idea is to apply the knowledge-transferred IV-exposure models to produce IV scores that explain additional exposure variation in the target population and thereby improve the efficiency of IV estimation.} However, the key challenge is that this information may not be fully transportable. As noted in the literature \citep{zhang2023new}, the SNP--exposure associations can differ substantially between different ancestry groups. Thus, external-source-based IV models may be useful for some exposures and target groups, but inaccurate for others. Naively using inaccurate external information can increase bias or variance relative to an estimator based only on the target data, a phenomenon commonly referred to as negative transfer.  In particular, when IV scores based on the source IV models are weakly associated with exposure in the target population or highly correlated with each other, directly using these IV scores to implement the standard two-stage least squares regression for MR analysis on the target may suffer from weak-IV-type bias and instability \citep{bound1995problems}.  
Therefore, knowledge transfer in MR requires methods that can adaptively borrow IV--exposure information while protecting against negative transfer.

Classical IV theory provides the foundation for causal inference
in the presence of unmeasured confounding \citep{anderson1950asymptotic,newey1990efficient,newey1993eff}. Related developments have addressed important complications such as high-dimensional instruments and weak instruments \citep{bound1995problems,caner2008nearly,newey2009generalized,wang2025debiased}. However, most of this literature focuses on IV analyses in a single population and does not directly address our IV transfer learning setting.  In our setting, the transferred IV scores may be weakly correlated with exposure due to the inaccuracy of external IV--exposure models or could be highly correlated with each other in the target population. There remains a key open problem on how to combine different IV scores in such a challenging setting, while simultaneously mitigating weak-IV bias and optimizing statistical efficiency.

Recent transfer learning approaches have focused mainly on improving prediction accuracy \citep{bastani2021predicting, li2022transfer, li2023estimation, tian2022transfer, cai2025semi}. Although an accurate prediction of the IV--exposure relationship is useful, it does not necessarily yield a robust and efficient causal effect estimator in the target population when the IV score based on the fitted model is used as an IV. 
Recently, several transfer learning methods have been developed in the context of MR \citep{hou2025temr,yang2025improving,hao2025transfer}. These works mainly focus on transferring causal effect estimates from a source population to a target population, typically relying on the assumption that the causal effects of exposures on outcomes are similar, though not necessarily identical, across the two populations. 
Methodologically, this line of work can be viewed as an extension of prediction-oriented transfer learning methods for model parameters \citep{li2022transfer,ruan2022improving} to the MR setting. In contrast, our framework exploits the specific structure of IV estimators and identifies IV strength as the bottleneck in studies of underrepresented subpopulations. Rather than directly transferring causal effects, our goal is to transfer the IV models to enhance the IV strength in the target population. This enables safe transfer estimation under weaker assumptions. Specifically, the causal effect and the IV-exposure relationship are both allowed to vary across populations, while knowledge transfer can still improve the robustness and efficiency of IV analysis. In a concurrent work, \cite{huang2026xmr} transfer the estimated linear coefficients of exposure against IVs using a parametric likelihood method based on a joint-normal assumption on the source and target coefficients. In contrast, our method can exploit nonlinear relationships between exposure and IVs without requiring the parametric model or joint-normal assumptions. Moreover, we establish rigorous statistical guarantees on weak-IV bias reduction, efficiency improvement, and negative-transfer protection.

\subsection{Our Contributions}

\begin{figure}[htb!]
    \centering
    \includegraphics[width=0.7\linewidth]{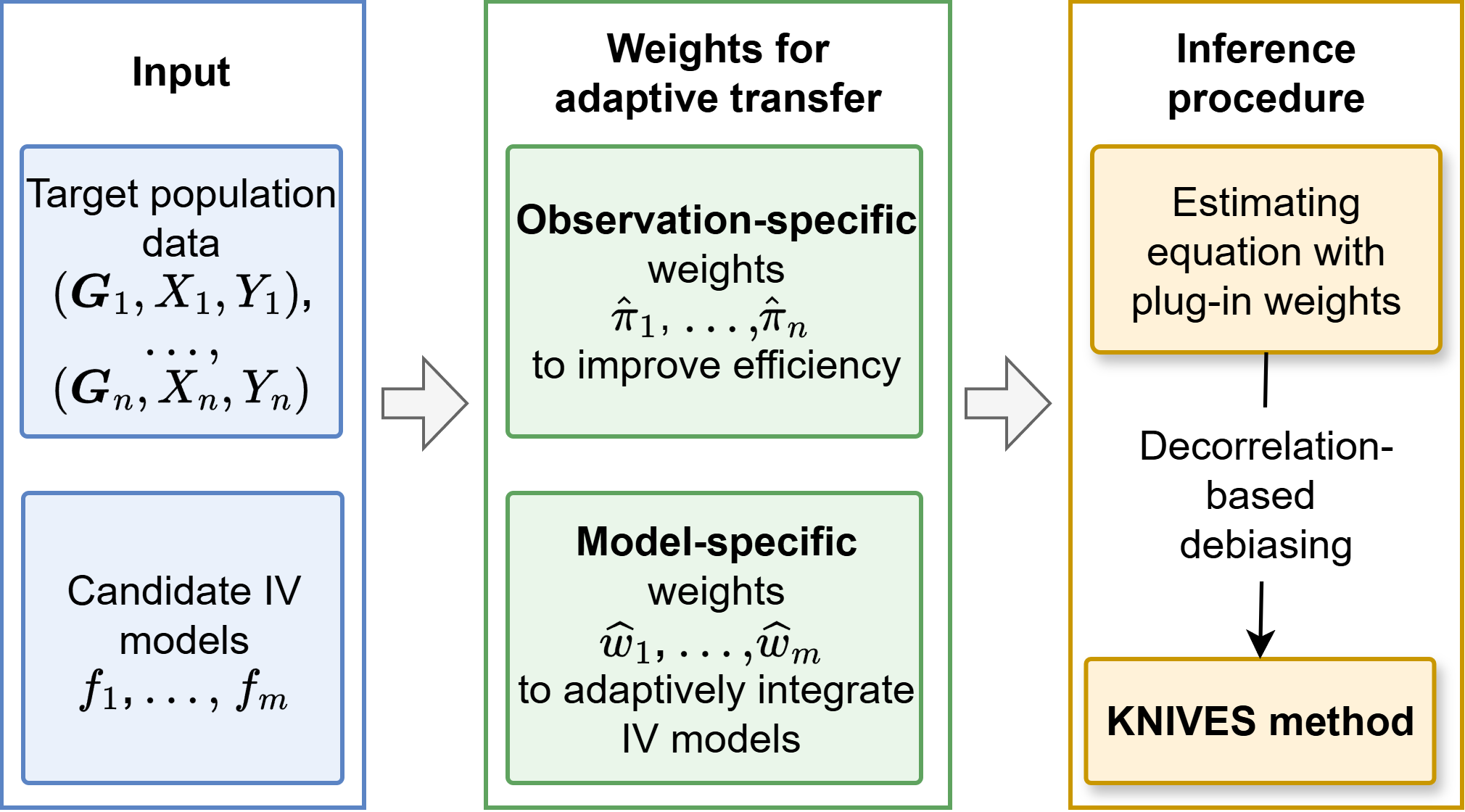}
    \caption{\label{fig:1} The Workflow of the KNowledge-powered IV Estimator for underrepresented Subpopulations (KNIVES).   The candidate IV models include source-pretrained models from external populations, target-trained models based solely on the target population, and transferred models adapted from source-pretrained models.}
\end{figure}

In this paper, we propose the KNowledge-powered IV Estimator for underrepresented Subpopulations (KNIVES), a robust transfer learning framework for improving IV estimation in underrepresented target populations. As illustrated in Figure \ref{fig:1}, KNIVES takes as input the target-population data and a collection of candidate IV--exposure models (``IV models'' for short) obtained from heterogeneous sources, including source-pretrained models, target-only models, and transfer learning models. KNIVES then constructs an aggregated IV score by linearly combining the candidate IV scores and further incorporates observation-specific weights to improve efficiency and accommodate heteroskedasticity in the target population.

 A central feature of KNIVES is that the transfer and aggregation of source models is not guided by prediction accuracy alone. Existing transfer learning methods select or combine source information using prediction-oriented criteria \citep{li2022transfer,tian2022transfer}. In IV analysis, however, a model predicting the exposure well does not necessarily give a stable and efficient causal effect estimator, especially when the resulting IV score is weakly associated with exposure or highly correlated with other candidate IV scores. Motivated by this distinction, KNIVES learns the model aggregation and observation-specific weights by optimizing an effective signal-to-noise ratio (SNR) criterion for the aggregated IV score. This criterion is directly tied to the downstream IV estimator in the sense that (i) it encourages a sufficiently strong aggregated IV score to avoid bias due to overly weak IV; and (ii) in the strong-IV regime, its reciprocal corresponds to the leading asymptotic variance of the causal effect estimator. Thus, the criterion provides an IV-specific principle for knowledge transfer, targeting robust and efficient causal estimation rather than exposure prediction alone.

Another key component of KNIVES is a decorrelation-based bias-correction procedure. A direct plug-in estimator that first computes and aggregates candidate IV scores and then uses the aggregated score in a standard IV estimating equation can still suffer from weak-IV-type bias, particularly when the candidate IV scores are weak or highly correlated with each other. KNIVES addresses this issue by modifying the estimating equation with a scalar decorrelation term. This correction complements the effective-SNR criterion in the sense that the criterion strengthens the aggregated IV score, while the decorrelation step controls the remaining weak-IV-type bias.

Our framework is flexible and can incorporate candidate IV models trained by a wide range of statistical or machine learning methods. Importantly, the final inference remains anchored to the target population. KNIVES allows both the IV--exposure relationship and the causal effect to differ between the source and target populations, and it does not require causal effects on source populations to be transportable to the target. It also allows the candidate IV models to be misspecified or inaccurate, because source-derived models enter only as candidates and are adaptively weighted using the target sample.


In summary, our contributions are threefold. First, we develop a transfer learning framework for IV analysis in underrepresented subpopulations and introduce an effective-SNR-based optimization criterion that is tailored to the downstream causal estimator rather than to prediction accuracy alone. Second, we develop a decorrelation procedure to correct the weak-IV-type bias induced by weak or highly correlated candidate IV scores. Third, we establish theoretical guarantees showing that KNIVES delivers valid inference under weak-IV regimes and achieves improved efficiency in strong-IV regimes, including protection against negative transfer. Numerical studies and two UK Biobank MR analyses further demonstrate the finite-sample performance of KNIVES.


\subsection{Organization of the Paper}
The rest of this paper is organized as follows. In Section \ref{sec: method}, we propose the KNIVES method and its cross-fitting version. Section \ref{sec: implementation} contains more details on practical implementation. In Section \ref{sec: theory}, we establish the theoretical properties of the KNIVES method. 
We report extensive simulation results and two real data analyses in Sections \ref{sec: simulation} and \ref{sec: real data}, followed by discussions. Additional simulations and all proofs are provided in the Supplementary Materials. 

\vspace{-0.1in}
\section{A Knowledge Transfer Framework for IV Estimation}\label{sec: method}

\subsection{The Limitations of IV Estimation Using Only the Data from the Target Underrepresented Subpopulations}


To illustrate the main idea, we focus on the regime without adjustment covariates and provide details for a more general setting with covariates in  Supplemental Materials. In the target population, suppose that the outcome $Y$ and exposure $X$ satisfy
\begin{equation}\label{eq: outcome model}
  Y = \beta_{0} + X\beta_* + \epsilon,
\end{equation}
where $\epsilon$ is a mean-zero error term and $\beta_*$ encodes the causal effect of $X$ on $Y$ of our interest. We allow $X$ to be correlated with $\epsilon$ due to unmeasured confounding effects. 
Consequently, directly regressing $Y$ against $X$ does not recover $\beta_*$. Assume that a vector $\bG$ of IVs is available and satisfies the following IV assumptions. (i) {Relevance: $X$ is associated with $\bG$, i.e., $\var\{E(X\mid \bG)\} \neq 0$;} (ii)  {Mean independence: The mean of $\epsilon$ is  independent of $\bG$, i.e., $E(\epsilon \mid \bG) = 0$;} (iii) {Exclusion restriction: $\bG$ does not directly affect $Y$, i.e., $\bG$ does not appear in the outcome model \eqref{eq: outcome model}.} 


According to the classic IV theory \citep{newey1993eff}, a semiparametric efficient estimator of the causal effect $\beta_*$ can be obtained by first transforming the IV vector into an optimal scalar variable and then separately regressing the outcome and the exposure against this variable.  In particular, let $f(\bG)$ denote a candidate scalar IV model mapping $\bG$ to a predictor of exposure $X$, termed an IV score in this paper. Under the IV assumptions (i)--(iii) introduced above, for any $f(\cdot)$, it holds that
\begin{equation}\label{eq: moment condition}
\begin{aligned}
   &\cov\{f(\bG), Y\} = \cov\{f(\bG), X\} \beta.\\
\end{aligned}
\end{equation}
Thus, we can estimate $\beta$ by 
$ \hbeta_{f} = \hcov\{f(\bG), Y\}/\hcov\{f(\bG), X\}$,
where $\hcov$ denotes the sample covariance on the target data. The properties of the causal estimator $\hbeta_{f}$ depend critically on the IV model $f(\cdot)$. For example, $\hbeta_{f}$ is inconsistent if $\cov\{f(\bG), X\}=0$. Conversely, assuming $\var\{E(X \mid \bG)\}\neq 0$ and a constant conditional error variance $\var(\epsilon\mid\bG)$, $\hbeta_{f}$ is semiparametrically efficient \citep{newey1993eff}, attaining the minimum asymptotic variance among a broad class of regular estimators, if $f(\cdot)$ is taken as the ideally optimal choice: $f_{\rm opt}(\bG) \propto E(X \mid \bG)$. 

In the scenario of target-data alone, one can first regress $X$ on $\bG$ to estimate $E(X\mid \bG)$, and then use the estimator denoted as $\tilde{f}_{\cT}(\bG)$ to construct $\hbeta_{\tilde{f}_{\cT}} =  \hcov\{\tilde{f}_{\cT}(\bG), Y\} / \hcov\{\tilde{f}_{\cT}(\bG), X\}$. This gives rise to the widely used two-stage least squares estimator (TSLS) \citep{anderson2005origins}.
The TSLS estimator is efficient when $\tilde{f}_{\cT}(\cdot)$ converges sufficiently well to $E(X \mid \bG)$. However, for underrepresented subpopulations, such as AFR or SAS individuals in the UK Biobank, $\tilde{f}_{\cT}(\cdot)$ is likely to be misspecified or estimated with substantial error because of the limited sample size and the potentially complex form of $E(X\mid \bG)$. This can substantially reduce the efficiency of $\hbeta_{\tilde{f}_{\cT}}$. Moreover, weak correlation between the IV score $\tilde{f}_{\cT}(\bG)$ and the exposure $X$ can lead to weak-IV bias in the conventional TSLS estimator \citep{bound1995problems}. On the other hand, given a IV model, such as $f_{\rm opt}$, the causal effect can be obtained by estimating only two covariance quantities, which is relatively straightforward even when the target-population sample size is limited. Therefore, the quality of the IV model constitutes a central bottleneck for IV estimation in underrepresented subpopulations.

To address this problem, our main idea is to leverage the knowledge from external source data to improve efficiency and mitigate weak IV bias. A natural strategy is to train a model for $E(X \mid \bG)$ using large-scale source data and use it as $f(\bG)$ to construct the TSLS estimator. However, due to the heterogeneity of $E(X \mid \bG)$ between the source and target, such a pretrained source model is not necessarily a better predictor in the target population compared to the target-only estimator, which may lead to a worse estimator of $\beta$. It is important to avoid such a negative transfer. Furthermore, multiple candidate models are often available, including pretrained source models, transfer learning models and target-data-only models, and their predictive performance on the target may vary substantially. A key challenge is therefore to determine how to effectively combine multiple candidate models while guarding against negative transfer. These considerations motivate the development of adaptive transfer learning methods for IV estimation and, in particular, the proposed KNIVES framework.

\subsection{IV Models Aggregation and the KNIVES Method}\label{subsec: aggregation}




The KNIVES construction has three steps. First, we obtain and apply a collection of candidate IV models to the target data and obtain multiple candidate IV
scores. Second, we adaptively combine these IV scores and assign observation-specific weights by optimizing an effective signal-to-noise ratio criterion, which is designed to avoid overly weak IV scores and to improve efficiency in the strong-IV regime. Third, we use the resulting weights and aggregated IV score in a bias-corrected estimating equation to obtain the final causal effect estimator.

\paragraph{Candidate IV models and aggregated IV score.} 
Let $f_1(\cdot),\ldots,f_m(\cdot)$ denote $m$ candidate IV--exposure models, each mapping
the IV vector to a scalar IV score that predicts exposure $X$. We treat these models as given in this section. In practice, these models can be trained using data from either the target or source population, based on parametric models or flexible black-box machine learning methods. The specific construction strategy for these models will be introduced in Section \ref{subsec: candidate set}. We
write $\fb(\bG) = \{f_{1}(\bG), \dots, f_{m}(\bG)\}^{\T}$ for the vector of candidate IV scores. Suppose we have $n$ i.i.d. observations $\sD_{n} = \{(Y_{i}, X_{i}, \bG_{i})\}_{i = 1}^{n}$ from the target population. Let $\bX = (X_{1}, \dots, X_{n})^{\T}$, $\bY = (Y_{1}, \dots, Y_{n})^{\T}$, and $\beps = (\epsilon_{1}, \dots, \epsilon_{n})^{\T}$ be the vectors of exposures, outcomes, and noise terms, respectively, where $\epsilon_{i} = Y_{i} - \beta_{0} - X_{i}\beta_*$. We collect the candidate IV scores into the $n\times m$ matrix $\bbF = \{\fb(\bG_{1}), \dots, \fb(\bG_{n})\}^{\T}$. We introduce a model weight vector $\bw=(w_1,\ldots,w_m)^\T$ and define the aggregated IV score for $\bG$ as
\[
\bw^\T \fb(\bG)=\sum_{j=1}^m w_j f_j(\bG).
\]
As will be introduced later, the weights in $\bw$ will be adaptively chosen using the target sample to produce a valid and efficient causal estimator.


\paragraph{Observation-specific working weights.}
In addition to assembling the candidate IV scores, KNIVES introduces  observation-specific weights in the target sample in the heteroskedastic setting where $\var(\epsilon\mid \bG)$ is allowed to depend on $\bG$. In specific, let $\pi(\bG;\btheta)>0$ be a parametric working weight function indexed by $\btheta$, and
assume that
$
\pi(\bG;\boldsymbol{0})\equiv 1.
$
Thus, $\btheta=\bzero$ corresponds to the unweighted case. The working weight $\pi(\bG;\btheta)>0$ allows different target observations to contribute differently to the causal estimator. 

\paragraph{The causal effect estimator with given $\bw$ and $\btheta$.}
For any $\btheta$, define the matrices
$\bbPi_{\btheta} = n^{-1}\diag\{\pi(\bG_{1}; \btheta), \dots, \pi(\bG_{n}; \btheta)\}$
and $\bbM_{\theta} = \bbI - \bone(\bone^{\T}\bbPi_{\btheta}\bone)^{-1}\bone^{\T}\bbPi_{\btheta}$, where $\mathbf{1}$ is the $n$-dimensional vector of ones. The matrix $\bbM_{\theta}$ is used to center
vectors with respect to the weighted empirical mean induced by $\pi(\bG_i;\btheta)$. Given the model weights in $\bw$ and the observation-specific weighting function $\pi(\cdot;\btheta)$, motivated by equation (\ref{eq: moment condition}), a natural plug-in estimator of $\beta_*$ can be constructed as
\[
\hbeta_{\bw, \btheta} = \frac{\hcov_{\btheta}\{\bw^{\T}\fb(\bG), Y\}}{\hcov_{\btheta}\{\bw^{\T}\fb(\bG), X\}} = 
\frac{\bw^{\T}\bbF^{\T}\bbPi_{\btheta}\bbM_{\btheta}\bY}{\bw^{\T}\bbF^{\T}\bbPi_{\btheta}\bbM_{\btheta}\bX},
\]
where $\hcov_{\btheta}$ is the empirical covariance operator on the target sample weighted by $\pi(\bG;\btheta)$ and $\bbPi_{\btheta}$ consists of the weights of the observations. This estimator is valid since it can be viewed as the TSLS estimator on some ``weighted'' population determined by $\pi(\bG;\btheta)$. It reduces to the standard TSLS estimator when $\btheta=\bzero$ so that $\pi(\bG;\btheta)\equiv 1$ and the aggregated IV score assigns weight $1$ to the IV model learned using only the target data and weight $0$ to all other candidate IV models.

\paragraph{A unified criterion for optimization.}
To reduce the bias and variance of $\hbeta_{\bw, \btheta}$, we next introduce an optimization criterion for the model weights $\bw$ and the observation-specific weight parameter $\btheta$. For any random variable $V$, let $E_{\btheta}(V) = E\{\pi(\bG;\btheta)V\}/E\{\pi(\bG; \btheta)\}$ denote its expectation on the target population weighted by the standardized version of the individual weight $\pi(\bG;\btheta)$. Also, let $\bgamma_{\btheta} = \cov_{\btheta}\{\fb(\bG), X\} = E(\pi(\bG; \btheta)[\fb(\bG) - E_{\btheta}\{\fb(\bG)\}]\{X - E_{\btheta}(X)\})$ denote the covariance between $\fb(\bG)$ and $X$ weighted by $\pi(\bG;\btheta)$, and
\[
  \bbSig_{\btheta} = E\left\{\pi(\bG; \btheta)^{2}\epsilon^{2}\left[\fb(\bG) - E_{\btheta}\{\fb(\bG)\}\right]\left[\fb(\bG) - E_{\btheta}\{\fb(\bG)\}\right]^{\T}\right\}
\]
be the weighted covariance matrix of $\fb(\bG)$. We define our key criterion function, the effective signal-to-noise ratio (SNR) as 
\[ 
\kappa(\bw,\btheta) =\frac{\left(\bw^{\T}\bgamma_{\btheta}\right)^{2}}{\bw^{\T}\bbSig_{\btheta} \bw}.
\]
The numerator of $\kappa(\bw,\btheta)$, equal to $[\cov_{\btheta}\{\bw^{\T}\fb(\bG), X\}]^{2}$, quantifies the signal strength of the aggregated IV score through its squared association with exposure $X$, whereas the denominator quantifies the noise level using the variability $\bw^{\T}\fb(\bG)$ and $\epsilon$.



According to our theoretical studies in Section \ref{sec: theory}, the effective SNR is crucial in characterizing the bias and variance of $\hbeta_{\bw, \btheta}$. First, it needs to be sufficiently large to avoid the situation of overly weak IV, where the estimator may suffer from non-negligible weak-IV-type bias. Second, when the aggregated IV score is strong enough to avoid weak-IV bias, the effective SNR is proportional to the reciprocal of the asymptotic variance of $\hbeta_{\bw, \btheta}$ that can be expressed as $\{n\kappa(\bw,\btheta)\}^{-1}$. Thus, we propose to minimize $\{\kappa(\bw,\btheta)\}^{-1}$ with respect to $\bw$ and $\btheta$, which protects the causal estimator $\hbeta_{\bw, \btheta}$ from excessive bias while minimizing its asymptotic variance in the strong IV regime. 


Note that for any given $\btheta$, the weight $\bw$ that minimizes $\{\kappa(\bw,\btheta)\}^{-1}$ has an explicit-form
$\bw_{\theta}:= \bbSig_{\btheta}^{+}\bgamma_{\btheta}$,
where $\bbA^{+}$ is the Moore-Penrose pseudo-inverse of a square matrix $\bbA$. Then, one can determine $\btheta$ by minimizing $\{\kappa(\bw_{\theta},\btheta)\}^{-1}=(\bgamma_{\btheta}^{\T}\bbSig_{\btheta}^{+}\bgamma_{\btheta})^{-1}$. Empirically, $\bgamma_{\btheta} = \cov_{\btheta}\{\fb(\bG), X\}$ can be estimated by $\hat\bgamma_{\btheta}:=\hcov_{\btheta}\{\fb(\bG), X\} = \bbF^{\T}\bbPi_{\btheta}\bbM_{\btheta}\bX$ and $\bbSig_{\btheta}$ can be estimated by:
\[
   \hat{\bbSig}_{\btheta} = n^{-1}\sum_{i = 1}^{n}\pi(\bG_{i}; \btheta)^{2}\hat{\epsilon}_{i}^{2}\left[\fb(\bG_{i}) - \hE_{\btheta}\{\fb(\bG)\}\right]\left[\fb(\bG_{i}) -  \hE_{\btheta}\{\fb(\bG)\}\right]^{\T},
\]
where $\hE_{\btheta}(\cdot)$ denotes the empirical version of the weighted expectation operator $E_{\btheta}(\cdot)$ and $\hat{\epsilon}_{i}$ is an estimate of the noise term $\epsilon_i$. To obtain $\hat{\epsilon}_{i}$, we first derive an initial estimator of $\beta_0$ and $\beta_*$ using the procedure described in the Supplementary Material. Then we can iteratively update $\hat{\epsilon}_{i}$ with the newly obtained estimators of $\beta_0$ and $\beta_*$ in each round until convergence.

Based on these, we can solve $\btheta$ from the regularized problem
\begin{equation}\label{eq: opt weight}
   \begin{aligned}
    \htheta=\arg\min_{\btheta} \left\{\left(\hat\bgamma_{\btheta}^{\T}\hat\bbSig_{\btheta}^{+}\hat\bgamma_{\btheta}\right)^{-1} + \lambda_{v} \|\btheta\|^{2}
       \right\},
   \end{aligned}
\end{equation}
where $\lambda_{v}$ is a tuning parameter and the $L_2$-penalty is added to mitigate the potential singularity issue of $\hat\bbSig_{\btheta}$ when $f_{1}(\bG), \dots, f_{m}(\bG)$ are weak or highly correlated with each other. In practice, we recommend choosing $\lambda_{v}$ to be large enough such that the objective function \eqref{eq: opt weight} is strongly convex in a neighborhood of zero. Based on $\htheta$, we can calculate the model weight estimator $\hat\bw:=\bw_{\htheta}=\hat\bbSig_{\hat\btheta}^{+}\hat\bgamma_{\hat\btheta}$ and the ``plug-in'' transfer learning estimator as $\hbeta_{\hat\bw, \hat\btheta} ={\hcov_{\hat\btheta}\{\hat\bw^{\T}\fb(\bG), Y\}}/{\hcov_{\hat\btheta}\{\hat\bw^{\T}\fb(\bG), X\}}$.



\begin{remark}[Role of the observation-specific weights]
When the aggregated IV model recovers the true $E(X\mid \bG)$ and the noise variance depends on $\bG$, our estimator achieves semiparametric efficiency if the observation-specific weight $\pi(G;\btheta)=\operatorname{var}(\epsilon\mid G)^{-1}$ \citep{hayashi2011econometrics}. However, in small target samples, this conditional variance is generally difficult to estimate reliably. To avoid unstable results, we recommend setting $\pi(\bG;\btheta)$ as a simple parametric working model. The weight $\pi(\bG;\btheta)$ can also improve efficiency when the candidate IV models are misspecified and the aggregated IV score $\bw^{\T}\fb(\bG)$ cannot approximate $E(X\mid \bG)$ well. This is realized by our data-driven weight that can downweight target observations on which the aggregated $\bw^{\T}\fb(\bG)$ is less accurate in terms of predicting $X$.
\end{remark}

\subsection{Bias of the Plug-in Estimation and Its Correction}\label{subsec: bias of plug-in}



Let the weighted IV-score aggregation matrix $\bbH_{\btheta} = \bbM_{\btheta}^{\T}\bbPi_{\btheta}\bbF\hat\bbSig_{\btheta}^{+}\bbF^{\T}\bbPi_{\btheta}\bbM_{\btheta}$. For notational simplicity, write $\bbPi_{\htheta}$, $\bbM_{\htheta}$, $\hat\bbSig_{\htheta}$ and $\bbH_{\htheta}$ as $\bbPi$, $\bbM$, $\bbSig$ and $\bbH$. The plug-in estimator $\hbeta_{\hat\bw, \hat\btheta}$ introduced in Section \ref{subsec: aggregation} can be biased when the candidate IV scores are weakly associated with exposure $X$ in the target population or highly correlated with each other. In this subsection, we identify the bias term induced by the correlation between exposure and the outcome error, and construct a decorrelated estimating equation to remove this bias.

To see the bias of the plug-in estimator, we note that $\hbeta_{\hat\bw, \hat\btheta}$ can be equivalently written as the solution to the estimating equation:
\begin{equation}\label{eq: plug-in}
  \bX^{\T}\bbH(\bY - \bX\beta) = 0,
\end{equation}
and evaluate its left-hand side at the true value $\beta_*$. By the definition of $\bbH$, we can derive that
$
\bX^{\T}\bbH(\bY - \bX\beta_*)=\bX^{\T}\bbH\beps,
$
where $\beps$ is the vector of the noise term defined in (\ref{eq: outcome model}). To focus on the main idea, we temporarily treat $\bbH$ as fixed. Then, we have
\begin{equation}
E\{\bX^{\T}\bbH(\bY - \bX\beta_*)\}
=
\sum_{i,j}h_{ij}E(X_i\epsilon_j)
=
\sum_{i=1}^n h_{ii}E(X_i\epsilon_i),
\label{eq: bias plug-in}
\end{equation}
where $h_{ij}$ is the $(i,j)$-th entry of $\bbH$. This derivation holds in the asymptotic sense, provided that $\bbH$ properly converges to some non-random quantities; see Section \ref{sec: theory}. The cross terms for $i\neq j$ in (\ref{eq: bias plug-in}) vanish because
different observations are independent and $E(\epsilon_i)=0$. However, the diagonal term $E(X_i\epsilon_i)$ generally does not vanish, because $X_i$ may be correlated with $\epsilon_i$ due to unmeasured confounding, which induces bias to equation (\ref{eq: plug-in}). This bias has a similar nature as the so-called weak IV bias \citep{bound1995problems} and is influential when $h_{ii}$ is large, which could be caused by a weak association of the candidate IV scores with $X$ or a high correlation between the IV scores. As illustrated by a numerical example in the Supplement, this weak-IV issue can make the plug-in estimator $\hbeta_{\hw, \htheta}$ severely biased.


Our key idea is to remove the weak-IV bias by decorrelating $\bX$ from $\beps$ in the plug-in estimating equation (\ref{eq: plug-in}). If the error vector $\beps$ were
observed, we could replace $\bX$ by $\bX-\bar\eta\beps$ in equation (\ref{eq: plug-in}) and construct the debiased estimating equation as:
\begin{equation}\label{eq: infeasible EE}
   (\bX - \bar\eta\beps)^{\T}\bbH(\bY - \bX\beta) = 0,    
\end{equation}
in which we choose $\bar\eta = E(\sum_{i = 1}^{n}h_{ii}X_{i}\epsilon_{i})/E(\sum_{i = 1}^{n}h_{ii}\epsilon_{i}^{2})$ so that the weighted diagonal covariance between $\bX-\bar\eta\beps$ and $\beps$ is zero. By encouraging the orthogonality between $\bX - \bar\eta\beps$ and $\beps$, our method can remove the ``diagonal'' component $h_{ii}E(X_i\epsilon_i)$ responsible for the bias of the plug-in estimating equation (\ref{eq: plug-in}).



Empirically, when the candidate value $\beta$ is close to the true $\beta_*$, $\beps$ used in equation (\ref{eq: infeasible EE}) can be well-approximated by $\bbM(\bY-\bX\beta)$ where the centering matrix $\bbM$ is used to remove the unknown intercept term. Then the coefficient $\bar\eta$ can be approximated by: 
\[
 \eta(\beta) = \frac{\bX^{\T}\bbM^{\T}\bbH_{\rm d}\bbM(\bY - \bX\beta)}{(\bY - \bX\beta)^{\T}\bbM^{\T}\bbH_{\rm d}\bbM(\bY - \bX\beta)}.
\]
Applying these approximations of $\beps$ and $\bar\eta$ to equation (\ref{eq: infeasible EE}) and using that $\bbH\bone=\bzero$, we obtain the KNIVES estimator $\hbeta_{\KNIVES}$ by solving the decorrelated estimating equation:
\begin{equation}\label{eq: pe}
   \{\bX- \eta(\beta) (\bY - \bX\beta)\}^{\T}\bbH(\bY - \bX\beta)= 0,
\end{equation}
where $\bbH_{\rm d} = \diag\{h_{11},\dots,h_{nn}\}$. This equation is one-dimensional and can be solved by standard root-finding algorithms in a neighborhood of the initial estimator described in Supplementary Material. The main steps for obtaining $\hbeta_{\KNIVES}$ are summarized in Algorithm \ref{alg: KNIVES}.


\begin{algorithm}[htb!!]
\renewcommand{\thealgorithm}{KNIVES}
\caption{\label{alg: KNIVES} KNIVES Estimation with Given Candidate IV Models.}
\label{alg:alg1}
\textbf{Input:} The target data $\sD_{n} = \{(Y_{i}, X_{i}, \bG_{i})\}_{i = 1}^{n}$ and a vector of candidate IV models $\fb$.
\begin{algorithmic}[1]
    \State{Obtain $\htheta$ by minimizing the regularized effective-SNR criterion in \eqref{eq: opt weight};}
   \State{Compute the matrix $\bbH_{\htheta}$ based on $\sD_{n}$, $\fb$ and $\htheta$, as described in Section \ref{subsec: bias of plug-in};}
	\State{Solve the estimating equation \eqref{eq: pe}.}
\end{algorithmic}
 \textbf{Output:} The solution of \eqref{eq: pe} denoted as $\hbeta_{\KNIVES}$.
\end{algorithm}

\subsection{The Cross-fitted KNIVES Method}\label{subsec: cross-fitting}

In Section \ref{subsec: aggregation}, we focus on the case when the candidate IV models are independent of the target sample $\sD_{n}$. However, as will be discussed in Section \ref{subsec: candidate set}, it is important to use the target data to calibrate external source IV models, which helps avoid negative knowledge transfer. In this situation, candidate IV models in $\fb$ depend on the target sample.

We use the cross-fitting strategy \citep{chernozhukov2018double} to avoid the overfit problem while properly accounting for the uncertainty arising from estimation of the IV models. This strategy works because equation \eqref{eq: moment condition} holds for any fixed $f$. Combined with the cross-fitting procedure, this property implies that estimation errors in the IV models contribute only higher-order terms to the resulting causal effect estimator as shown in the proof of Theorem \ref{thm: KNIVES CF AN}. Specifically, we first divide $\sD_{n}$ into $K$ disjoint subsets $\sD_{n}^{(1)}, \dots, \sD_{n}^{(K)}$ with equal size for some integer $K > 0$. Let $\sD_{n}^{(-k)}$ be the dataset that consists of all observations in $\sD_{n}$ except for those in $\sD_{n}^{(k)}$. Then for $k = 1,\dots, K$, we obtain the vector of candidate IV models $\tf^{(k)}$ based on $\sD_{n}^{(-k)}$ using the method described in Section \ref{subsec: candidate set} and apply Algorithm \ref{alg: KNIVES} to $\sD_{n}^{(k)}$ and $\tf^{(k)}$. Denote the output estimator as $\tbeta_{\KNIVES}^{(k)}$. The cross-fitted KNIVES estimator is
\begin{equation*}
   \bar{\beta}_{\KNIVES} = \frac{1}{K}\sum_{k = 1}^{K}\tbeta_{\KNIVES}^{(k)}.
\end{equation*}
This strategy preserves the validity of KNIVES while allowing the candidate set to include flexible target-trained or transferred IV models introduced in Section \ref{subsec: candidate set}. For each fold $k$, models $\tilde f^{(k)}$ are trained on $\sD_{n}^{(-k)}$ and, thus, independent of the held-out observations $\sD_{n}^{(k)}$ used in the estimating equation for $\beta$. This reduces overfitting bias caused by reusing the same observations for both model training and causal estimation.

\section{Practical Implementation}\label{sec: implementation}
\subsection{Construction of the Candidate Set for IV Models}\label{subsec: candidate set}


Our KNIVES procedure takes a collection of candidate IV--exposure models as input. We now describe how these models can be constructed in practice. Again, to keep the notation simple, we present the case without observed adjustment covariates. Let $\mathcal F=\{f_{\bxi}:\bxi\in\bbXi\}$ be a user-specified model class, where \(f_{\bxi}(\bG)\) maps the IV vector \(\bG\) to a scalar IV score predicting exposure \(X\), and \(\bxi\) denotes the model parameter vector. The class \(\mathcal F\) may be linear, as in \(f_{\bxi}(\bG)=(1, \bG^\T)\bxi\), or nonlinear, e.g., a neural network or a tree model. 

Let \(\ell(x,u)\) be a loss function for predicting \(x\) by \(u\). For continuous exposures considered in our numerical studies, we use the squared loss \(\ell(x,u)=(x-u)^2\). For binary exposures, We can use the entropy loss. Let \(\mathcal P_{\lambda}(\bxi)\) be a penalty for the model parameter \(\bxi\), with the tuning parameter \(\lambda\); common examples include \(\lambda\|\bxi\|_1\) and \(\lambda\|\bxi\|_2^2\). For each fold \(k=1,\ldots,K\), recall that \(\sD_n^{(k)}\) and \(\sD_n^{(-k)}\) denote the held-out and training sets of the target sample. The candidate IV models \(\tf^{(k)}\) are constructed from three types of learners: source-pretrained learners, target-only learners, and transfer learners.

\paragraph{Source-pretrained learners.} Suppose that \(S\) source populations are available, and for each $s\in\{1,\ldots,S\}$, $\{(X_i^{(s)},\bG_i^{(s)}):~i=1,\ldots,N_s\}$ are the IV--exposure data from the \(s\)-th source. A source-pretrained learner is obtained by fitting the IV--exposure model: 
\[ 
\hat \bxi_{\cS}^{(s)} = \arg\min_{\bxi\in\bbXi} \left\{ \frac{1}{N_s}\sum_{i=1}^{N_s} \ell\{X_i^{(s)},f_{\bxi}(\bG_i^{(s)})\} + \mathcal P_{\lambda_s}(\bxi) \right\}, \qquad \hat  f_{\cS}^{(s)}=f_{\hat  \bxi_{\cS}^{(s)}}. 
\] 

This training objective is used to illustrate the common constructions of candidate IV models between the source and target. It is worth noticing that only fitted IV scores enter the aggregation step and the theoretical development in Section \ref{sec: theory}, which is not necessarily based on parametric models. Thus, our framework actually accommodates generic machine learning algorithms, including random forests, gradient boosting, and neural networks, and does not require a likelihood-based loss or an explicit parameter penalty. For methods like random forest and neural networks, regularization may instead be implemented through tree depth or early stopping.

Because \(\hat f_{\cS}^{(s)}\) is trained outside the target sample, the same source-pretrained learner can be included in \(\tf^{(k)}\) for every fold \(k\). This learner can have low variance because it is trained on large source data, but it may be poorly aligned with the target population when the IV--exposure relationship differs substantially across populations.

\paragraph{Target-only learners.} The target-only learner uses only the target training observations: 
\[ 
\tilde \bxi_{\cT}^{(k)} = \arg\min_{\bxi\in\bbXi} \left\{ \frac{1}{|\cI_{n}^{(-k)}|} {\sum}_{i\in \cI_{n}^{(-k)}} \ell\{X_i,f_{\bxi}(\bG_i)\} + \mathcal P_{\lambda_{\cT}}(\bxi) \right\}, \qquad \tilde f_{\cT}^{(k)}=f_{\tilde \bxi_{\cT}^{(k)}},
\] 
where $\cI_{n}^{(k)}$ and $\cI_{n}^{(-k)}$ are the index sets of $\sD_{n}^{(k)}$ and $\sD_{n}^{(-k)}$, respectively. This learner is directly targeted to the population of interest and free from bias due to distributional heterogeneity. However, when the target population is underrepresented, the target sample size may be too small to estimate a strong IV--exposure model reliably. In this situation, including \(\tilde f_{\cT}^{(k)}\) in the candidate set is still important because it provides a target-data-only benchmark, allowing KNIVES to avoid negative transfer when the source-pretrained models are not informative.

\paragraph{Transfer learners.} The third type of learner uses the source-pretrained model as prior knowledge and adapts it using the target training sample. Given a source-pretrained parameter \(\hat  \bxi_{\cS}^{(s)}\), we obtain the fine-tuned parameter by 
\[ 
\tilde \bxi_{\FT}^{(k,s)} = \arg\min_{\bxi\in\bbXi} \left\{ \frac{1}{|\cI_{n}^{(-k)}|} {\sum}_{i\in \cI_{n}^{(-k)}} \ell\{X_i,f_{\bxi}(\bG_i)\} + \mathcal P_{\lambda_{\FT}}\big(\bxi-\hat  \bxi_{\cS}^{(s)}\big) \right\}, \ \text{and}\ \tilde f_{\FT}^{(k,s)} = f_{\tilde \bxi_{\FT}^{(k,s)}}. 
\] 
Here, \(\mathcal P_{\lambda_{\FT}}\big(\bxi-\hat  \bxi_{\cS}^{(s)}\big)\) penalizes the deviation of $\bxi$ from the source-pretrained parameter. For example, one may use \(\lambda_{\FT}\|\bxi-\hat  \bxi_{\cS}^{(s)}\|_1\) for sparse linear models \citeg{li2022transfer} or \(\lambda_{\FT}\|\bxi-\hat  \bxi_{\cS}^{(s)}\|_2^2\) for ridge-type shrinkage \citeg{bastani2021predicting}. For neural networks, this corresponds to initializing the model at the source-pretrained weights and regularizing the updated weights toward the pretrained weights. This learner provides a compromise between source pretraining and target-only learning in the sense that it borrows strength from the source population while still adapting the IV--exposure model to the target data.

For each fold \(k\), we collect all learners in 
$
\tf^{(k)} = (\hat f_{\cS}^{(1)},\ldots,\hat f_{\cS}^{(S)}, \tilde f_{\cT}^{(k)}, \tilde f_{\FT}^{(k,1)},\ldots, \tilde f_{\FT}^{(k,S)} )^\T, 
$
and set it as the candidate IV models used in Section \ref{sec: method}. In Mendelian randomization, published polygenic scores can be included as pretrained candidate IV models as well. 
The tuning parameters in the training objectives above can be selected by cross-validation or IV models obtained under multiple choices of tuning parameters can be directly included as separate candidates. The three learners have complementary roles. Source-pretrained learners can be strong when the source and target IV--exposure relationships are shared between the source and target. Target-only learners are free from population heterogeneity but may be unstable in small target samples. Transfer learners interpolate between these two extremes. 

\subsection{IV Strength Evaluation}\label{subsec: IV strength}

The strength of the IV plays an important role in determining the performance of the IV estimators, especially in underrepresented subpopulations where the IVs can be weak or unstable. In practice, it can provide a summary measure of how informative the candidate IV models are in explaining the variation in exposure $X$. In our framework, stronger IV scores correspond to a larger covariance between $X$ and the aggregated IV score introduced in Section \ref{sec: method}, relative to the residual variation of $X$. This motivates us to define the population IV strength measure as
\[
   r_{*} = \frac{n\bgamma_{*}^{\T}\var_{\btheta_{*}}\{\fb(\bG)\}^{+}\bgamma_{*}}{\var_{\btheta_{*}}(X) - \bgamma_{*}^{\T}\var_{\btheta_{*}}\{\fb(\bG)\}^{+}\bgamma_{*}},
\]
where $\bgamma_{*} = \cov_{\btheta_{*}}\{\fb(\bG), X\}$, \(\var_{\btheta_{*}}(\cdot)\) is the variance on target population weighted by \(\pi(\bG;\btheta_{*})\), and
\begin{equation}\label{eq: pop var}
    \btheta_{*} = \mathop{\arg\min}_{\btheta}\left\{(\bgamma_{\btheta}^{\T}\bbSig_{\btheta}^{+}\bgamma_{\btheta})^{-1}  + \lambda_{v} \|\btheta\|^{2}\right\}
\end{equation}
is the population counterpart of $\htheta$. 
The numerator in $r_{*}$ reflects the variation in \(X\) explained by the IV score, while  the denominator captures the remaining unexplained variation. This is analogous to the classical concentration parameter in standard IV analysis \citep{stock2002survey}, but adapted to our transfer learning setting where the true conditional mean \(E(X \mid \bG)\) may be difficult to reliably estimate in the target population. In addition, the IV strength $r_{*}$ is scaled by $n$. Thus, under a sufficiently strong IV–exposure association, it is of the same order as the sample size $n$. Hence, $r_{*}$ can be interpreted as an ``effective sample size'' induced by the aggregated IV score which ranges from zero to infinity. Larger values of $r_{*}$ indicate greater IV information and, consequently, tend to support more reliable IV inference. However, there is no universal cutoff for $r_{*}$ that guarantees reliable inference. Even in classical IV settings, the threshold for diagnosing weak instruments can depend on the number of instruments and the desired degree of control over estimation bias or test size distortion \citep{stock2002survey}. As a practical rule of thumb, following the conventional weak-instrument guidance introduced in \citet{stock2002survey}, the asymptotic approximation may be regarded as reasonably accurate when $r_{*}\geq 10$.

In practice, \(r_{*}\) is unknown and is estimated by plugging in empirical counterparts of \(\gamma_{*}\) and the weighted covariance matrices. The resulting estimator is denoted by \(\hat r\). For the case with cross-fitting, let $\hat{r}^{(k)}$ be the estimated IV strength based on $\sD_{n}^{(k)}$ and $\tf^{(k)}$. The cross-fitted IV strength estimator can be estimated by their average $K^{-1}\sum_{k=1}^{K}\hat{r}^{(k)}$. This IV strength measure is primarily used as a diagnostic tool to assess whether our derived IV score contains sufficient information for reliable causal estimation and to compare the effectiveness of different model construction strategies in the transfer learning setting. 

\section{Theoretical Results}\label{sec: theory}

Let $\|\cdot\|$ denote the Euclidean norm for vectors and the spectral norm 
for matrices. Let $\kappa_{*} = \bgamma_{*}^{\T}\bbSig_{*}^{+}\bgamma_{*}$ be the population version of the effective SNR for the optimal model- and observation-specific weights. Throughout the theoretical analysis, we assume that the number of folds in cross-fitting $K$ is fixed while the number of candidate IV models $m$ and the effective SNR $\kappa_{*}$ are allowed to depend on the target sample size $n$, although we suppress the 
$n$-subscript for notational convenience. 

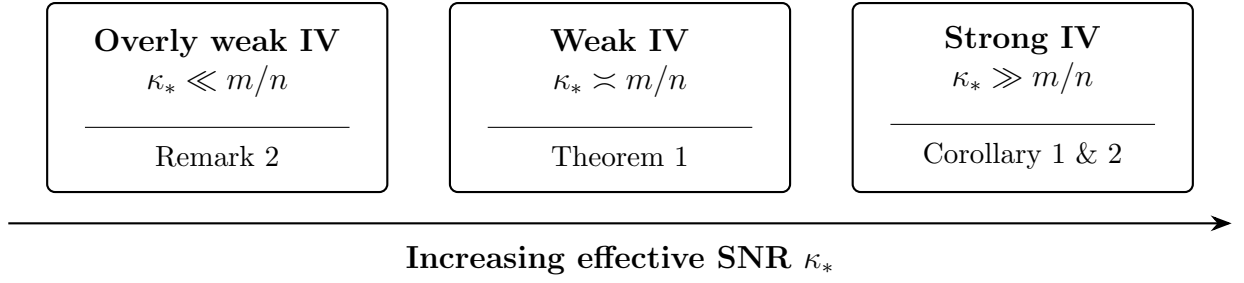
\begin{figure}[htbp]
    \centering
    \begin{tikzpicture}[
        node distance=0.5cm and 0.8cm,
        box/.style={
            draw, 
            rectangle, 
            rounded corners=3pt, 
            minimum width=4.5cm, 
            minimum height=2.5cm, 
            align=center, 
            thick
        },
        arrow/.style={-{Stealth[scale=1.2]}, thick}
    ]

    \node[box] (weakest) {
        \textbf{Overly weak IV} \\
        $\kappa_* \ll m/n$ \\
        \makebox[3.5cm]{\hrulefill} \\
        \small Remark \ref{rem:over:weak}
    };

    \node[box, right=of weakest] (weak) {
        \textbf{Weak IV} \\
        $\kappa_* \asymp m/n$ \\
        \makebox[3.5cm]{\hrulefill} \\
        \small Theorem \ref{thm: KNIVES CF AN}
    };

    \node[box, right=of weak] (strong) {
        \textbf{Strong IV} \\
        $\kappa_* \gg m/n$ \\
        \makebox[3.5cm]{\hrulefill} \\
        \small Corollary \ref{cor: strong} \&  \ref{cor: no negative trans} 
    };

    \draw[arrow] 
        ([yshift=-0.4cm, xshift=-0.5cm]weakest.south west) -- 
        ([yshift=-0.4cm, xshift=0.5cm]strong.south east) 
        node[midway, below=0.15cm] {\textbf{Increasing effective SNR $\kappa_*$}};


    \end{tikzpicture}
    \caption{An illustration of the property of KNIVES under different regimes of effective SNRs in the high-dimensional setting where $m\to \infty$.}
    \label{fig: eff IV strength}
\end{figure}

We establish the theoretical properties of KNIVES in three different regimes of $\kappa_*$, as outlined in Figure~\ref{fig: eff IV strength}. First, in the overly weak IV regime with $\kappa_* \ll m/n$, the instrumental signal is too weak to guarantee asymptotic unbiasedness; see Remark \ref{rem:over:weak}. Second, in the weak IV regime with $\kappa_{*} \asymp m/n$, the IV signal is moderate but still susceptible to second-order bias, making our decorrelation-based debiasing step crucial for achieving valid inference; see Theorem \ref{thm: KNIVES CF AN}. Finally, in the strong IV regime with $\kappa_* \gg m/n$, the instruments are sufficiently informative that the weak-IV-related terms become asymptotically negligible. In this scenario, KNIVES avoids negative transfer and can achieve semiparametric efficiency when the nuisance models are correctly specified; see Corollaries \ref{cor: strong} and \ref{cor: no negative trans}.  



Our studies rely on Conditions S1 to S5 in the Supplementary Material. Conditions S1 and S3 impose standard bounded moment and non-degeneracy assumptions on the variables. Condition S2 restricts the growth rate of the number of candidate IV models $m$ and the nuisance estimation errors. These rate restrictions are standard in the weak-IV literature. They ensure that the estimation error of multiple candidate models does not overwhelm the IV signal. We also allow the components of the candidate model vector $\fb$ to be highly correlated, in the sense that $\|\bbSig_*^{+}\|$ may diverge at appropriate rates depending on $m$ and $n$. Condition S4 adapts the above requirements to the cross-fitting framework. Condition S5 requires that the learned candidate IV models obtained for each fold $k$ converge to the same deterministic function $\fb$ in a proper sense. In general, these are widely used and reasonable assumptions in the literature on IV and semiparametric inference \citep{newey2009generalized,chernozhukov2018double}. 

In Theorem \ref{thm: KNIVES CF AN}, we focus on the asymptotic normality of the cross-fitted KNIVES estimator $\bar{\beta}_{\KNIVES}$. In Theorem S1 of the Supplement, we present the theoretical properties of the KNIVES estimator when candidate IV models are pretrained using data independent of the target sample $\sD_{n}$ so cross-fitting is omitted. The consistency of our estimators is justified in and Proposition~S1 in the Supplementary 
Material.


\begin{theorem}\label{thm: KNIVES CF AN}
     Under Conditions S1 -- S5 in the Supplementary Material, if $n\kappa_{*}/m$ is bounded away from zero and $n\kappa_{*} \to \infty$, we have
    \begin{equation}\label{eq: KNIVES CF AN general}
        \sqrt{n}\sigma_{*}^{-1}\left(\bar{\beta}_{\KNIVES} - \beta_*\right) \to N(0, 1),
    \end{equation}
    where
    $
      \sigma_{*}^{2} = \kappa_{*}^{-1} + Kn^{-1}\kappa_{*}^{-2}\delta_{\KNIVES},
    $  and $\delta_{\KNIVES}$ is the variance of the normalized second order term in the asymptotic expansion of $\bar{\beta}_{\KNIVES} - \beta_*$ whose explicit formula is provided the Supplementary Material.
\end{theorem}

\begin{remark}
\label{rem:over:weak}
The effective SNR $\kappa_{*}$ plays a crucial role in Theorem~\ref{thm: KNIVES CF AN}. The asymptotic unbiasedness of $\bar{\beta}_{\KNIVES}$ requires that the effective SNR is not {overly weak}, as imposed that
$\kappa_{*}\gtrsim m/n$. Otherwise, higher-order terms in the expansion of $\bar{\beta}_{\KNIVES} - \beta_*$ could induce excessive bias to our estimator. Also, the asymptotic variance $\sigma_{*}^{2}$ decreases as $\kappa_{*}$ increases. Given the limited sample size of the target data, $\kappa_{*}$ can be increased by incorporating informative pretrained or transfer IV models obtained leveraging the rich source samples. In addition, introducing the observation-specific weight $\pi(\bG;\btheta_{*})$ can further increase the effective SNR, because the population loss function in \eqref{eq: pop var} favors values of $\btheta$ associated with larger $\kappa_*=\bgamma_{\btheta_*}^{\T}\bbSig_{*}^{+}\bgamma_{\btheta_*}$. This highlights the benefits of knowledge transfer and individual weighting. 
\end{remark}

\begin{remark}\label{remark: eta and r}
    The parameters $\kappa_*$ and $r_{*}$ defined in Section \ref{subsec: IV strength} are both related to the overall strength of the IV models. However, they serve distinct purposes in our analysis. The parameter $r_*$ depends only on the distribution of $X$ and $\bG$ and is primarily intended as a diagnostic for the appropriateness of using the IV, which is analogous to the concentration parameter in standard IV analysis \citep{stock2002survey}. In contrast, $\kappa_{*}$ involves $Y$ and is more directly related to asymptotic properties of the KNIVES estimator.
\end{remark}

We next provide a more detailed explanation on the asymptotic behavior of $\bar{\beta}_{\KNIVES}$. Let $\eta_{*}$ be the population counterpart of $\bar{\eta}$, whose explicit expression is given in Supplementary Material. Define $\fb_{*}(\bG_{i}) = \fb(\bG_{i}) - E_{\btheta_{*}}\{\fb(\bG)\}$, $X_{i}^{*} = X_{i} - E_{\btheta_{*}}(X)$, and $Y_{i}^{*} = Y_{i} - E_{\btheta_{*}}(Y)$. Under the conditions of Theorem~\ref{thm: KNIVES CF AN}, the variance terms in the asymptotic expansion of $\bar{\beta}_{\KNIVES} - \beta_*$ are given by
\begin{equation}\label{eq: KNIVES expansion cf}
    \begin{aligned}
&L_{n} + U_{n}\\ &\coloneqq \left(n\kappa_{*}\right)^{-1}\sum_{i = 1}^{n} \pi(\bG_{i}; \btheta_{*})\bgamma_{*}^{\T}\bbSig_{*}^{+}\fb_{*}(G_{i})\epsilon_{i}\\
&
\quad + K(n^{2}\kappa_{*})^{-1}\sum_{k = 1}^{K}\sum_{\substack{i,j \in \cI_{n}^{(k)}\\i\neq j}}\pi(\bG_{j}; \btheta_{*})\{\pi(\bG_{i}; \btheta_{*})(X_{i}^{*} - \eta_{*}\epsilon_{i})\fb_{*}(G_{i}) - \bgamma_{*}\}^{\T}\bbSig_{*}^{+}\fb_{*}(G_{j})\epsilon_{j}.
\end{aligned}
\end{equation}
In (\ref{eq: KNIVES expansion cf}), $L_n$ is a mean-zero term obtained by averaging the first-order terms in the asymptotic expansion of $\tbeta_{\KNIVES}^{(k)}-\beta_{*}$ for $k = 1,\dots, K$.
The term $U_n$ is an incomplete $U$-statistic arising from the second-order expansion of $\tbeta_{\KNIVES}^{(k)}-\beta_{*}$. It is mean-zero and asymptotically normal due to the decorrelation-based debiasing step removing the second-order bias in $\bar{\beta}_{\KNIVES}$. Without this debiasing step, the second-order bias may dominate the asymptotic expansion and overwhelm the benefits of transfer learning when $\kappa_{*}$ is small. In practice, $\sigma_{*}^{2}$ can be estimated by using its explicit expression and replacing the unknown nuisance quantities with their empirical estimators.

Note that $\var(n^{1/2}L_{n}) = \kappa_{*}^{-1}$ and  $\var(n^{1/2}U_{n}) = Kn^{-1}\kappa_{*}^{-2}\delta_{\KNIVES}$ correspond to the first and second terms in $\sigma_{*}^{2} = \kappa_{*}^{-1} + Kn^{-1}\kappa_{*}^{-2}\delta_{\KNIVES}$.
In addition, we have $\var(n^{1/2}U_{n})/\var(n^{1/2}L_{n}) = O(m/(n\kappa_{*}))$ since the quantity $\delta_{\KNIVES} = O(m)$ under Condition S1. Based on these, our estimator has distinct asymptotic behaviors in the weak IV and strong IV regimes defined in Figure \ref{fig: eff IV strength}. First, in the {\em strong-IV regime} with $\kappa_{*}\gg m/n$ and $\var(n^{1/2}U_{n})/\var(n^{1/2}L_{n}) \to 0$, $L_n$ is the only leading term that contributes to the asymptotic variance of $\bar{\beta}_{\KNIVES}$. Second, in the {\em weak-IV regime} with $\kappa_* \asymp m/n$, the influence of the second-order term $U_n$ becomes non-negligible. Note that $\var(n^{1/2}U_{n})$ has a factor $K$, the number of folds for cross-fitting. This is because the data-splitting and averaging procedures cause $U_n$ to be an incomplete $U$-statistic with a variance $K$ times greater than that of the corresponding complete $U$-statistic. Therefore, to control the variance in the weak-IV regime, we suggest using a relatively small number of folds such as $K=2$ or $K=3$.


In the following Corollaries \ref{cor: strong} and \ref{cor: no negative trans}, we study the efficiency properties of $\bar{\beta}_{\KNIVES}$ derived by knowledge transfer, in the strong-IV regime when $n\kappa_{*}\gg m$ and our estimator is asymptotically normal with variance $(n\kappa_{*})^{-1}$.
\begin{corollary}\label{cor: strong}
    Under the conditions of Theorem \ref{thm: KNIVES CF AN}, if we further assume $n \kappa_{*}/m \to \infty$, then
    \begin{equation*}
        \left(n\kappa_{*}\right)^{1/2}\left(\bar{\beta}_{\KNIVES} - \beta_*\right) \to N(0, 1).
    \end{equation*}
    The asymptotic variance of $\bar{\beta}_{\KNIVES}$ achieves the semiparametric efficiency bound if $E\{E(X\mid \bG)^2\} > 0$, the conditional mean $E(X\mid \bG)$ can be expressed as a linear combination of the candidate IV models and the weight $\pi(\cdot;\btheta_*)$ is correctly specified up to a constant in the sense that $E[\{c\pi(\bG;\btheta_{*}) - \var(\epsilon\mid \bG)^{-1}\}^{2}] \to 0$ for some constant $c>0$.
\end{corollary}

\begin{remark}\label{remark: intrinsic eff}
In the strong-IV regime, our estimator still enjoys favorable efficiency properties even when candidate IV models do not recover $E(X\mid \bG)$ and $\bar{\beta}_{\KNIVES}$ does not attain semiparametric efficiency. To see this, we note that for any $\bw$ satisfying $\sqrt{n}\bw^{\T}\bgamma_{\bzero}\to\infty$, the asymptotic variance of $\hbeta_{\bw^{\T}\fb}$ is
$
  {n^{-1}\bw^{\T}\bbSig_{\bzero}^{+}\bw}/{(\bw^{\T}\bgamma_{\bzero})^{2}}.
$
Under the conditions of Corollary~\ref{cor: strong}, the asymptotic variance of $\bar{\beta}_{\KNIVES}$ is $(n\kappa_{*})^{-1}$, where
\[
\begin{aligned}
     &\left(n\kappa_{*}\right)^{-1}
  =
  \left(n\bgamma_{*}^{\T}\bbSig_{*}^{+}\bgamma_{*}\right)^{-1}
  \leq n^{-1}\min_{\btheta}\left\{(\bgamma_{\btheta}^{\T}\bbSig_{\btheta}^{+}\bgamma_{\btheta})^{-1}  + \lambda_{v} \|\btheta\|^{2}\right\}\\
  &\leq n^{-1}\{(\bgamma_{\bzero}^{\T}\bbSig_{\bzero}^{+}\bgamma_{\bzero})^{-1} + \lambda_{v} \|\bzero\|^{2}\}
  =
  \min_{\bw}
  \left\{
     \frac{n^{-1}\bw^{\T}\bbSig_{\bzero}\bw}
          {(\bw^{\T}\bgamma_{\bzero})^{2}}
  \right\}.
\end{aligned}
\]
Here, the second inequality follows from equation (\ref{eq: pop var}). The inequality highlights the benefit of individualized weights in improving efficiency compared to the unweighted estimator. The right-hand side of the above inequality is precisely the minimum asymptotic variance attainable by the class of estimators $\hbeta_{\bw^{\T}\fb}$ over different choices of $\bw$. Let ${\rm avar}$ denote the asymptotic variance of an asymptotically normal estimator. Then we have
\[
  {\rm avar}(\bar{\beta}_{\KNIVES})
  \leq
  {\rm avar}(\hbeta_{\bw^{\T}\fb})
\]
for any $\bw$ satisfying $\sqrt{n}\bw^{\T}\bgamma_{\bzero}\to\infty$. In particular, this result implies that $\bar{\beta}_{\KNIVES}$ is at least as efficient as the single-model estimator $\hbeta_{f_j}$ for each $j=1,\ldots,m$.
\end{remark}
KNIVES is more efficient than the estimator based on any single candidate IV model under regularity conditions, as discussed in Remark~\ref{remark: intrinsic eff}. Therefore, it can avoid negative transfer when the target-data-only IV model $\tilde f_{\cT}^{(k)}$ is included as a candidate. Let $f_{\cT}$ be the probability limit of $\tilde f_{\cT}^{(k)}$. We introduce the following corollary to justify the protection against negative transfer in our framework. 

\begin{corollary}\label{cor: no negative trans}
  Suppose $\sqrt{n}\cov(f_{\cT}(\bG), X) \to \infty$ and the target population-based estimator $\hbeta_{\tilde{f}_{\cT}}$ is asymptotically normal and has the same asymptotic distribution as $\hbeta_{f_{\cT}}$. When $\fb$ contains $f_{\cT}$ as its component and the conditions of Corollary \ref{cor: strong} hold, we have ${\rm avar}(\bar{\beta}_{\KNIVES})  \leq {\rm avar}(\hbeta_{\tilde{f}_{\cT}})$.
\end{corollary}

In Supplement, we provide theoretical comparisons between the decorrelation-based debiasing approach used in KNIVES and alternative weak-IV-bias-correction strategies from the single-population IV literature that could be adapted to our setting, including diagonal removal \citep{mikusheva2024weak} and generalized empirical likelihood (GEL) \citep{newey2009generalized,caner2012cue}. Compared with diagonal removal, our approach not only eliminates the bias but also improves efficiency of the resulting estimator through projection the of $X$ to $\epsilon$. In particular, Corollary~S1 shows that our decorrelation-based estimator is strictly more efficient than the diagonal removal estimator. Moreover, compared to GEL, our approach is more stable in the presence of highly correlated candidate IV models. These comparisons demonstrate the advantages of KNIVES in terms of both statistical efficiency and stability.

\section{Simulation Studies}\label{sec: simulation}

We conducted simulation studies to evaluate two key components of KNIVES: adaptive aggregation of candidate IV models through the effective-SNR criterion and decorrelation-based bias correction. To keep the main text focused, we provide full implementation details, including coefficient values, genotype generation, tuning parameters, and additional numerical results, in Supplementary Material. Throughout the simulations, we implement the cross-fitted KNIVES estimator with $K=3$ folds and use
$
\pi(\bG;\btheta)=\exp\{\btheta^\T \fb(\bG)\}
$
as the working individual weight model. Each simulation is repeated 200 times. For evaluation, we report absolute bias, empirical standard error (SE), and root mean squared error (RMSE), unless otherwise stated. The specific IV models adopted in different simulation settings are introduced in Sections \ref{subsec: sim linear} and \ref{subsec: sim nonlinear}, respectively.

We consider two representative regimes. The first is a weak-IV setting, where the IVs are high-dimensional and weakly associated with exposure $X$ in the target population. This setting is designed to evaluate both the benefit of aggregating multiple candidate IV models and the ability of the decorrelation step to control
weak-IV bias. The second is a strong-IV setting, where the candidate IV scores are sufficiently informative for exposure $X$ on target. In this regime, weak-IV bias is less prominent, and the main question is whether KNIVES improves efficiency by adaptively combining candidate models. 

\subsection{Weak-IV Setting: Benefits of Transfer and Decorrelation}\label{subsec: sim linear}

We first consider a weak-IV setting based on a linear IV--exposure model that mimics real-world MR studies targeting underrepresented populations. Briefly, the target sample size is $1{,}050$, and each source population has sample size $20{,}000$. The IV vector $\bG$ is generated through HAPGEN2 \citep{su2011hapgen2} based on the 1000 Genomes Project (phase 3) \citep{10002015global} with an African reference panel. Given the computational intensity required to generate genotype data via Hapgen2, we utilized the existing $120,000$ subjects dataset from the previous study as a pool for our simulation runs. Exposure $X$ follows a linear model that involves observed covariates, IVs, and an unmeasured confounder. Three source populations are generated with varying degrees of heterogeneity from the target population in terms of the IV--exposure relationship. The level of heterogeneity is indexed by $h\in\{0,0.25,0.5,0.75,1\}$, with a larger $h$ indicating a greater discrepancy between the source and the target. Detailed data generating mechanisms are given in the Supplementary Material.

We construct candidate IV models using the three strategies described in Section \ref{subsec: candidate set}, including target-only learning, source-pretraining, and source-to-target fine-tuning. In this high-dimensional and weak IV setting, all candidate models are fitted using Lasso. To demonstrate the benefit of model aggregation, we compare KNIVES with estimators based on each single candidate IV model. For each single-model estimator, we apply the same decorrelation-based estimating equation as KNIVES but use only one candidate IV score. Hence, these single-model benchmarks benefit from decorrelation but do not leverage multiple candidate models.

\begin{figure}[htb!]
    \centering
    \includegraphics[width=0.9\linewidth]{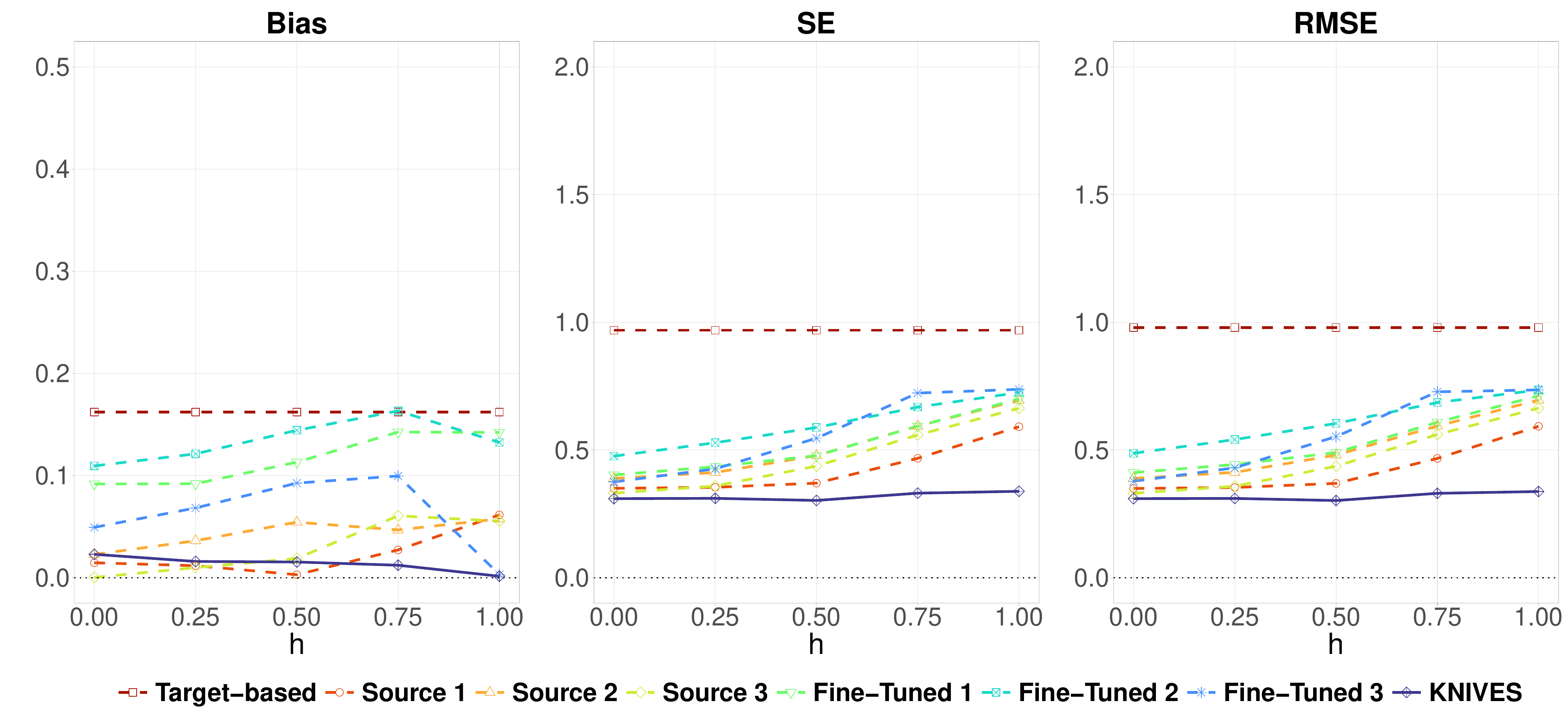}
    \caption{\label{fig: linear-sg} Performance of single-model estimators and KNIVES in the weak-IV setting. The parameter $h\in\{0,0.25,0.5,0.75,1\}$ controls the degree of heterogeneity on IV--exposure relationships, with larger $h$ indicating greater source--target discrepancy. ``Target-based'' denotes the IV model trained only on the target training data. ``Source $s$'' denotes the source-pretrained IV model from the $s$-th source. ``Fine-Tuned $s$'' denotes the transfer IV model obtained by fine-tuning the pretrained model from the $s$-th source using the target data. The panels show absolute bias, empirical standard error (SE), and root mean squared error (RMSE) across 200 simulation replications. Points are means and error bars indicate standard error over the replications.}
\end{figure}

Figure~\ref{fig: linear-sg} summarizes the performance of the single-model estimators and KNIVES. The target-only estimator has large bias and large SE, reflecting the instability of estimating the IV--exposure relationship using only the small target sample. For example, when $h=1$, the target-only estimator has absolute bias around $0.16$ and SE around $0.97$. Source-based and fine-tuned estimators improve over the target-only estimator in some cases, but their performance becomes less stable as the source--target heterogeneity increases; at $h=1$, their SEs are roughly between $0.60$ and $0.73$. In contrast, KNIVES maintains a near-zero bias and SE around $0.34$ at $h=1$, reducing the SE by around 65\% relative to the target-only estimator and by around 45\% relative to most source-based and fine-tuned single-model estimators. A similar pattern appears in RMSE. At $h=1$, KNIVES has RMSE around $0.34$, significantly lower than $0.98$ for the target-only estimator and $0.60$--$0.75$ for the source-based and fine-tuned estimators. These single-model benchmarks already use the same decorrelation correction as KNIVES, so the additional improvement of KNIVES reflects the benefit of adaptive aggregation across multiple candidate IV models.

We next evaluate the decorrelation component of KNIVES by comparing it with direct extensions of existing methods for weak or highly correlated IVs. Specifically, we compare KNIVES with standard TSLS, the jackknife IV estimator (JIV)\citep{angrist1999jackknife}, which is a diagonal removal-type approach, and GEL \citep{newey2009generalized,caner2012cue}, all implemented with cross-fitting using the same collection of candidate IV scores. This comparison focuses on whether an estimator can control the bias when candidate IV scores are weak and highly correlated. 

\begin{figure}[htb!]
    \centering
   \includegraphics[width=0.9\linewidth]{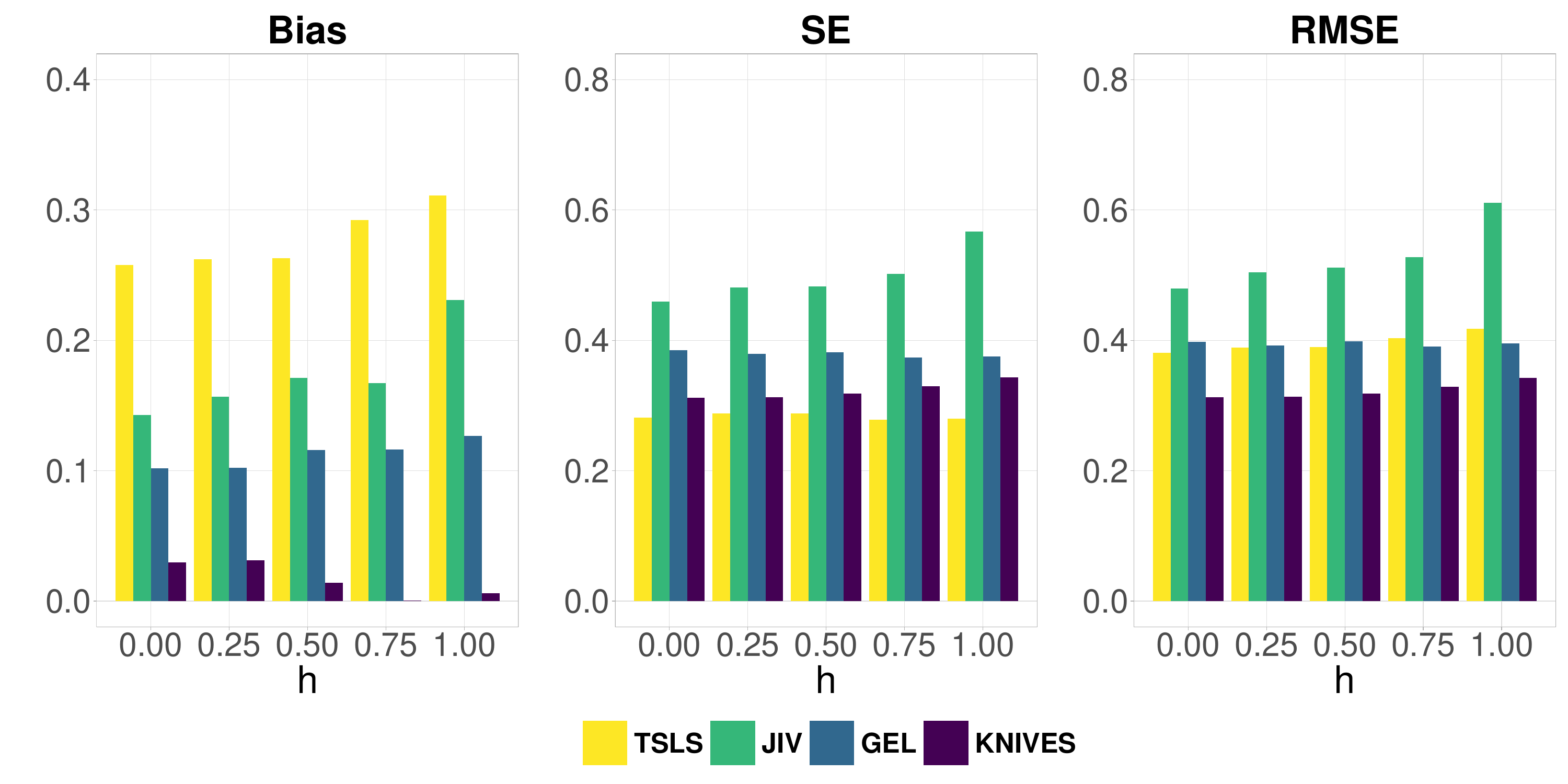}
    \caption{\label{fig: linear-Comb} Comparison of KNIVES with existing combination methods in the weak-IV setting. The parameter $h\in\{0,0.25,0.5,0.75,1\}$ indexes the degree of source--target heterogeneity in the IV--exposure relationship. The panels report absolute bias, empirical standard error (SE), and root mean squared error (RMSE) across 200 simulation replications.}
    
\end{figure}

Figure~\ref{fig: linear-Comb} shows that TSLS suffers from substantial bias, consistent with its known sensitivity to weak instruments. When $h=1$, the absolute bias is about $0.23$ for JIV, and about $0.13$ for GEL, whereas the absolute bias of KNIVES is close to zero. The same pattern is already visible when $h=0$: the absolute biases are approximately $0.25$, $0.15$, and $0.10$ for TSLS, JIV and GEL, respectively, compared to approximately $0.03$ for KNIVES. Although JIV and GEL partially reduce weak-IV bias relative to TSLS, they remain affected by the weakness and high correlation of the candidate IV scores. KNIVES achieves the smallest bias and the lowest RMSE across all values of $h$. These results demonstrate that our decorrelation step is important for controlling weak-IV-type bias in the transfer learning setting.

\subsection{Strong-IV Setting: Efficiency Gains from Aggregation}\label{subsec: sim nonlinear}

We next consider a relatively strong-IV setting in which candidate IV scores are substantially associated with exposure $X$. This setting is designed to evaluate the
efficiency gain from adaptive aggregation when weak-IV bias is less prominent. Exposure is generated from a nonlinear function of the IVs and observed covariates, and the result $Y$ depends linearly on exposure $X$, covariates, and the unmeasured confounder. The target sample size is again $1{,}050$, and the source sample size is $20{,}000$. Full details of the data generation process are provided in the Supplementary Material. For this setting, we construct a richer candidate set using several learning algorithms. Source-pretrained models are fitted using Lasso, neural networks (NN), random forests (RF), and gradient boosting (GB). Target-only models are fitted using Lasso, RF, and GB and we drop the target-only NN model because it is very unstable. Fine-tuned models are constructed using Lasso, NN, RF, and GB by adapting source-trained models to the target training sample. KNIVES then adaptively aggregates all candidate IV models.

\begin{figure}[htb!]
    \centering
    \includegraphics[width=0.8\linewidth]{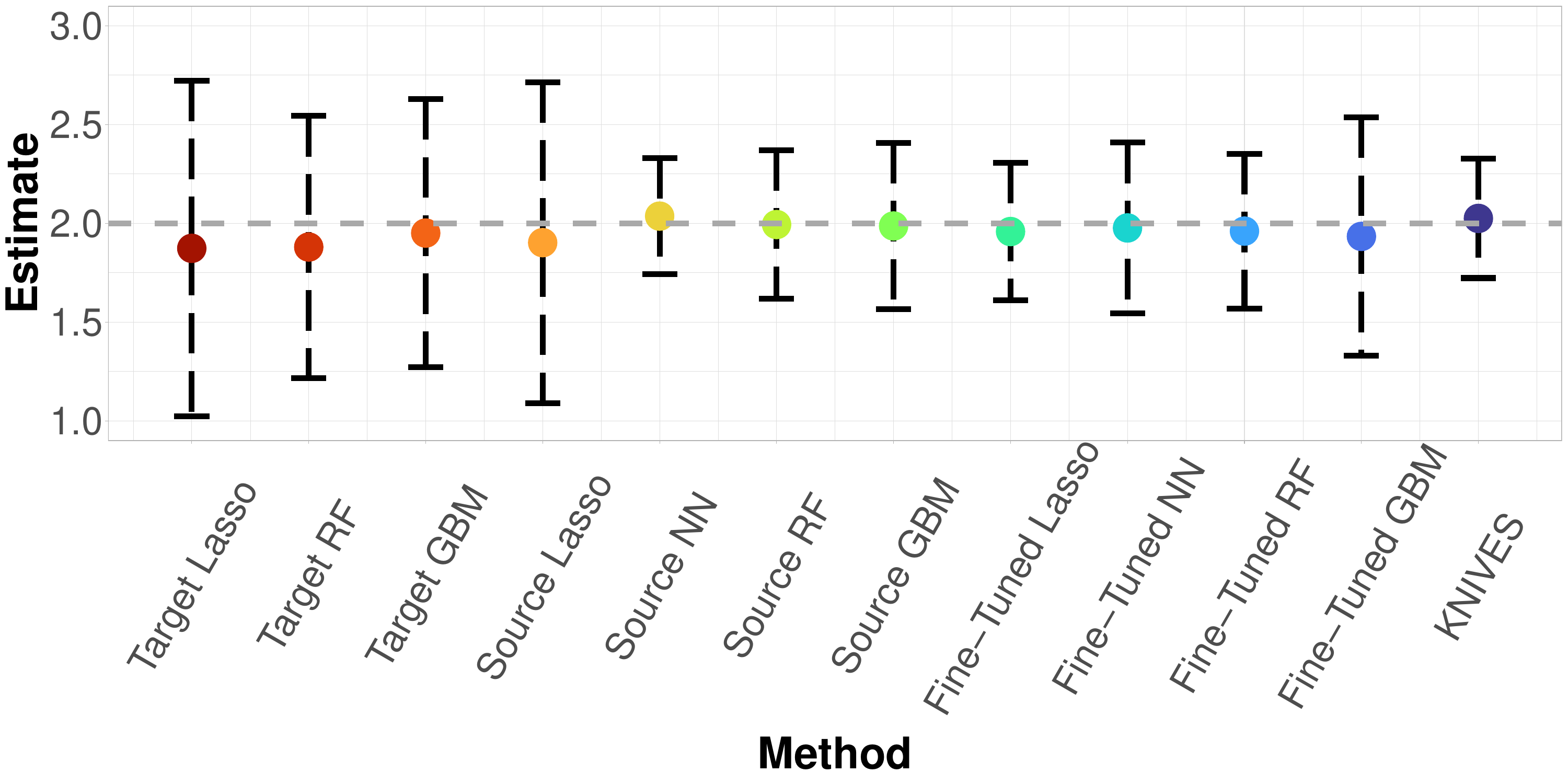}
    \caption{Performance of single-model estimators and KNIVES in the strong-IV setting with a nonlinear IV--exposure model. The candidate set includes source-pretrained models fitted by Lasso, neural networks (NN), random forests (RF), and gradient boosting (GB); target-only models fitted by Lasso, RF, and GB; and fine-tuned models fitted by Lasso, NN, RF, and GB. Points denote the mean causal effect estimates across 200 simulation replications, and error bars denote $\pm$ one empirical standard error. The dashed horizontal line indicates the true causal effect $\beta_*=2$.}
    \label{fig: nonlinear-sg}
\end{figure}

Figure~\ref{fig: nonlinear-sg} reports the mean and SE of the causal effect estimators. All estimators have relatively small biases, which confirms that weak-IV bias is not the main difficulty in this setting. The main difference lies in variability. The target-only estimators have large SEs because the target sample is small and the nonlinear IV--exposure relationship is difficult to estimate using target data alone; for example, the SE is about $0.84$ for Target Lasso and about $0.65$--$0.70$ for Target RF and Target GB. Some source-pretrained and fine-tuned models are more efficient than target-only models, with the best single-model estimator having SE around $0.30$--$0.35$. KNIVES achieves an SE of approximately $0.30$, matching or improving the most efficient single candidate. Thus, in the strong-IV regime, the primary advantage of KNIVES is variance reduction through adaptive aggregation rather than bias reduction.

In the strong-IV setting, additional comparisons with TSLS, JIV, and GEL using the same collection of candidate IV scores are reported in the Supplementary Material. In this regime, most methods have small bias, but KNIVES remains the most efficient because it effectively combines information across all candidate IV models while stabilizing the effect of redundant candidate scores.

\section{Real-World Study}\label{sec: real data}
\subsection{Effects of LDL on CAD in Underrepresented Populations}\label{subsec: LDL&CAD}

Quantifying the causal effect of low-density lipoprotein cholesterol (LDL-C) on coronary artery disease (CAD) is a central question in cardiovascular epidemiology. Mendelian randomization (MR) uses genetic variants as instrumental variables to address unmeasured confounding in observational studies. However, reliable IV estimation in underrepresented ancestry subpopulations remains challenging due to limited sample sizes and weak instrument strength. We perform an IV analysis to study the causal effect of LDL-C on CAD in African (AFR) and South Asian (SAS) populations using data from the UK Biobank. Although the data comprises approximately $500,000$ participants in total, the sample sizes for AFR and SAS populations are relatively small (around $7,000$ each). Thus, knowledge transfer from other populations, e.g., European (EUR) with a sample size greater than $420,000$, is essential for reliable and efficient IV analysis in underrepresented subpopulations. The EUR and SAS populations are used as source populations when inferring the causal effect in AFR. Similarly, EUR and AFR are used as sources when targeting SAS.

We conducted LD-clumping to select nearly independent SNPs based on the reference sample that matches the target population from the 1000 Genomes Project (phase 3) and the $p$-values from the same population as the target population in the ancestry-specific GWAS for LDL by the Global Lipids Genetics Consortium \citep{graham2021power}. Then, we select SNPs with $p$-values no larger than $10^{-2}$ as candidate IVs. This leads to $1240$- and $1312$-dimensional $\bG$ when the target population is AFR and SAS, respectively. Lasso is employed to fit the candidate IV models using the three approaches discussed in Section \ref{subsec: candidate set}. The resulting models are denoted by combinations of the training approach and the source population. For example, ``Source EUR'' denotes the source-based model trained on EUR and ``Fine-Tuned EUR'' denotes the model obtained by fine-tuning the pretrained EUR model using the target data. The cross-fitted KNIVES estimator is implemented with three folds. The causal effect estimation results are summarized in Table \ref{table: LDL-CAD}.

Table \ref{table: LDL-CAD} shows that KNIVES improves both precision and IV strength when estimating the effect of LDL-C on CAD in the two target populations. In AFR, all methods yield positive effect estimates and several reach nominal significance. KNIVES achieves the largest estimated IV strength, with $\hat r=320$, compared with $250$ for the target-only model and $136$--$269$ for the source-pretrained and fine-tuned models. It also gives the smallest SE, $0.011$, compared with $0.012$--$0.018$ for the single-model estimators. The KNIVES estimate is $0.029$ with 95\% CI $[0.007,0.052]$ and $p$-value $0.009$, which supports a positive causal effect of LDL-C on CAD in the AFR target population.

The gain from knowledge transfer is more pronounced in SAS. The target-only estimator has weak IV strength, with $\hat r=36.6$, and is imprecise, with SE $0.074$. The EUR-based source and fine-tuned models improve the IV strength ($\hat r=75.1$ and $79.1$), but their confidence intervals still include zero. KNIVES further increases the IV strength to $\hat r=97.2$, about 2.7 times that of the target-only estimator, and reduces the SE to $0.042$. Consequently, KNIVES is the only method that detects a significant positive effect of LDL-C on CAD in SAS at the 0.05 level, with a point estimate $0.085$, 95\% CI $[0.002,0.168]$ and $p$-value $0.044$. These results suggest that KNIVES can effectively borrow external IV--exposure information while preserving target-population inference, especially when the target-only IV score is weak.

\begin{table}[htb!]
    \centering
    \begin{tabular}{clccccc}
    \toprule
    Ancestry & Method & Estimate & SE & $p$-value & CI & $\hat{r}$\\
    \midrule
    \multirow{6}{*}{AFR} 
    & Target     & 0.035 & 0.013 & 0.010 & $[0.008, 0.061]$ & 250\\
    & Source EUR & 0.026 & 0.014 & 0.054 & $[0.000, 0.053]$ & 221\\
    & Fine-Tuned EUR  & 0.026 & 0.012 & 0.034 & $[0.002, 0.051]$ & 269\\
    & Source SAS & 0.043 & 0.018 & 0.018 & $[0.007, 0.078]$ & 136\\
    & Fine-Tuned SAS  & 0.035 & 0.013 & 0.007 & $[0.009, 0.061]$ & 254\\
    & KNIVES     & 0.029 & 0.011 & 0.009 & $[0.007, 0.052]$ & 320\\
    \midrule
    \multirow{6}{*}{SAS} 
    & Target     & 0.074 & 0.074 & 0.316 & $[-0.071, 0.220]$ & 36.6\\
    & Source EUR & 0.067 & 0.046 & 0.142 & $[-0.023, 0.157]$ & 75.1\\
    & Fine-Tuned EUR  & 0.063 & 0.048 & 0.195 & $[-0.032, 0.158]$ & 79.1\\
    & Source AFR & 0.049 & 0.067 & 0.466 & $[-0.082, 0.179]$ & 33.7\\
    & Fine-Tuned AFR  & 0.061 & 0.091 & 0.503 & $[-0.118, 0.240]$ & 36.4\\
    & KNIVES     & 0.085 & 0.042 & 0.044 & $[0.002, 0.168]$ & 97.2\\
    \bottomrule
\end{tabular}
\caption{\label{table: LDL-CAD} Estimated causal effect of low-density lipoprotein cholesterol (LDL-C) on coronary artery
disease (CAD) for African-ancestry (AFR) and South Asian-ancestry (SAS) target populations. Reported are the point estimate, standard error (SE), $p$-value, 95\% confidence interval (CI), and estimated IV strength $\hat r$ defined in Section \ref{subsec: IV strength}. ``Source'' denotes a source-pretrained IV model, and ``Fine-Tuned'' denotes a transfer IV model. EUR, AFR, and SAS denote European, African, and South Asian ancestry, respectively.}
\end{table}

\subsection{Effects of BMI on SBP in Underrepresented Populations}

Blood pressure is a common risk factor for cardiovascular disease and stroke, and body mass index (BMI) is known to be positively associated with blood pressure in observational studies \citep{mokdad2003prevalence}. Understanding whether this association reflects a causal relationship is of substantial public health importance and MR provides a natural framework for this question. However, as in the LDL–CAD analysis, inference in underrepresented ancestry groups is limited by weak instruments and small sample sizes. We perform an IV analysis of the causal effect of BMI on systolic blood pressure (SBP) in AFR and SAS using UK Biobank data. We follow the same IV construction pipeline as in Section \ref{subsec: LDL&CAD} (including LD-clumping and ancestry-specific SNP selection) and therefore omit the technical details here. The same construction strategy is used for candidate IV models to ensure comparability between analyses.

Table \ref{table: BMI-SBP} shows that the BMI--SBP analysis is more challenging because several candidate IV scores are extremely weak in the target populations. In AFR, the target-only model has an IV strength $\hat r=0.065$, and SAS-based source and fine-tuned models have $\hat r=1.60$ and $0.328$, respectively. Recall that the IV strength measure $\hat r$ is scaled by the target sample size $n$ to be comparable to the effective sample size in classical IV analysis. Therefore, values such as $\hat r =0.3$ indicate extremely weak effective instrument strength, corresponding to very limited usable information relative to the sample size. Among the usable single-model estimators (with $\hat r>5$), Source EUR and Fine-Tuned EUR have moderate IV strength ($\hat r=23.5$ and $22.6$) but remain imprecise, with SEs $0.708$ and $0.822$. By combining the information about the IV-exposure relationship in the target-only, source-pretrained, and fine-tuned IV models, KNIVES increases the IV strength to $\hat r=30.5$ and gives the smallest SE, $0.659$, with a point estimate $0.869$, 95\% CI $[-0.423,2.16]$, and $p$-value $0.187$. Although not statistically significant, KNIVES provides the strongest and most precise estimate among the methods under comparison.

When targeting SAS, the target-only estimator has limited IV strength, with $\hat r=10.1$, and produces an unstable estimate of $-1.06$ with SE $1.64$. The EUR-based source and fine-tuned models substantially improve IV strength and give positive estimates with SEs around $0.41$. KNIVES further increases the IV strength to $\hat r=108$ and achieves the smallest SE, $0.396$, with a point estimate $0.633$ and $p$-value $0.109$. Thus, although the BMI--SBP analysis does not provide significant evidence at the 0.05 level, KNIVES improves IV strength and precision in both underrepresented subpopulations and produces the most stable estimates.

\begin{table}[htb!]
    \centering
    \begin{tabular}{clccccc}
    \toprule
    Ancestry & Method & Estimate & SE & $p$-value & CI & $\hat{r}$\\
    \midrule
    \multirow{6}{*}{AFR} 
    & Target     & ---   & ---   & ---   & ---                 & 0.065\\
    & Source EUR & 0.574 & 0.708 & 0.418 & $[-0.814, 1.96]$   & 23.5\\
    & Fine-Tuned EUR  & 0.467 & 0.822 & 0.570 & $[-1.14, 2.08]$    & 22.6\\
    & Source SAS & ---   & ---   & ---   & ---                 & 1.60\\
    & Fine-Tuned SAS  & ---   & ---   & ---   & ---                 & 0.328\\
    & KNIVES     & 0.869 & 0.659 & 0.187 & $[-0.423, 2.16]$   & 30.5\\
    \midrule
    \multirow{6}{*}{SAS} 
    & Target     & -1.06 & 1.64  & 0.517 & $[-4.27, 2.15]$    & 10.1\\
    & Source EUR & 0.576 & 0.412 & 0.163 & $[-0.232, 1.38]$   & 98.5\\
    & Fine-Tuned EUR  & 0.568 & 0.418 & 0.174 & $[-0.251, 1.39]$   & 99.3\\
    & Source AFR & ---   & ---   & ---   & ---                 & 0.668\\
    & Fine-Tuned AFR  & -0.761& 1.88  & 0.686 & $[-4.45, 2.93]$    & 7.18\\
    & KNIVES     & 0.633 & 0.396 & 0.109 & $[-0.142, 1.41]$   & 108\\
    \bottomrule
\end{tabular}
 \caption{\label{table: BMI-SBP} Estimated causal effect of body mass index (BMI) on systolic blood pressure (SBP) for African-ancestry (AFR) and South Asian-ancestry (SAS) target populations. Reported are the point estimate, standard error (SE), $p$-value, 95\% confidence interval (CI), and estimated IV strength $\hat r$ defined in Section \ref{subsec: IV strength}. ``Source'' denotes a source-pretrained IV model, and ``Fine-Tuned'' denotes a transfer IV model. EUR, AFR, and SAS denote European, African, and South Asian ancestry, respectively. Entries are shown as ``---'' when $\hat r<5$, indicating that the corresponding IV score is too weak for stable causal effect estimation.}
\end{table}

\section{Discussion}\label{sec: discuss}

We propose KNIVES, a transfer learning framework for IV analysis in underrepresented subpopulations. It constructs an aggregated IV score by combining multiple candidate IV--exposure models from heterogeneous sources and improves efficiency via observation-specific weighting. A key feature of KNIVES is that its transfer step is guided by an effective-SNR criterion linked to IV strength and causal estimation efficiency, rather than by exposure prediction accuracy alone. To address weak-IV-type bias, we introduce a decorrelation approach for bias correction. Together, these components enable efficient and stable IV estimation. KNIVES uses external information without imposing homogeneity between populations. Both the IV--exposure relationship and the causal effect are allowed to vary, and source models serve only as auxiliary information. This avoids negative transfer while enhancing the IV estimation when external models are informative. Theoretical results show that KNIVES is asymptotically normal and more efficient than estimators based on any single candidate IV model, while remaining robust to weak and highly correlated IV scores.

Despite these advantages, our current framework is based on individual-level data, which may not always be available in large-scale genetic studies. Extending KNIVES to a fully summary-statistics-based implementation is an important direction for future work. In addition, we currently assume a classical exclusion restriction. More general settings allowing direct effects of instruments on the outcome are discussed in the Supplementary Material, where we consider a balanced direct-effect structure under which the moment condition is preserved in expectation. In addition, developing methods that is robust under directional pleiotropy or correlated invalid IVs remains an important open problem.


{\setstretch{0.8} 
\bibliographystyle{apalike}
\bibliography{TransIV}

\begin{thebibliography}{}

\bibitem[Anderson, 2005]{anderson2005origins}
Anderson, T.~W. (2005).
\newblock Origins of the limited information maximum likelihood and two-stage least squares estimators.
\newblock {\em Journal of Econometrics}, 127(1):1--16.

\bibitem[Anderson and Rubin, 1950]{anderson1950asymptotic}
Anderson, T.~W. and Rubin, H. (1950).
\newblock The asymptotic properties of estimates of the parameters of a single equation in a complete system of stochastic equations.
\newblock {\em The Annals of Mathematical Statistics}, 21(4):570--582.

\bibitem[Angrist et~al., 1999]{angrist1999jackknife}
Angrist, J.~D., Imbens, G.~W., and Krueger, A.~B. (1999).
\newblock Jackknife instrumental variables estimation.
\newblock {\em Journal of Applied Econometrics}, 14(1):57--67.

\bibitem[Bastani, 2021]{bastani2021predicting}
Bastani, H. (2021).
\newblock Predicting with proxies: Transfer learning in high dimension.
\newblock {\em Management Science}, 67(5):2964--2984.

\bibitem[Bound et~al., 1995]{bound1995problems}
Bound, J., Jaeger, D.~A., and Baker, R.~M. (1995).
\newblock Problems with instrumental variables estimation when the correlation between the instruments and the endogenous explanatory variable is weak.
\newblock {\em Journal of the American Statistical Association}, 90(430):443--450.

\bibitem[Burgess et~al., 2019]{burgess2019guidelines}
Burgess, S., Smith, G.~D., Davies, N.~M., Dudbridge, F., Gill, D., Glymour, M.~M., Hartwig, F.~P., Holmes, M.~V., Minelli, C., Relton, C.~L., et~al. (2019).
\newblock Guidelines for performing mendelian randomization investigations.
\newblock {\em Wellcome Open Research}, 4.

\bibitem[Cai et~al., 2025]{cai2025semi}
Cai, T., Li, M., and Liu, M. (2025).
\newblock Semi-supervised triply robust inductive transfer learning.
\newblock {\em Journal of the American Statistical Association}, 120(550):1037--1047.

\bibitem[Caner, 2008]{caner2008nearly}
Caner, M. (2008).
\newblock Nearly-singular design in gmm and generalized empirical likelihood estimators.
\newblock {\em Journal of Econometrics}, 144(2):511--523.

\bibitem[Caner and Y{\i}ld{\i}z, 2012]{caner2012cue}
Caner, M. and Y{\i}ld{\i}z, N. (2012).
\newblock Cue with many weak instruments and nearly singular design.
\newblock {\em Journal of Econometrics}, 170(2):422--441.

\bibitem[Chernozhukov et~al., 2018]{chernozhukov2018double}
Chernozhukov, V., Chetverikov, D., Demirer, M., Duflo, E., Hansen, C., Newey, W., and Robins, J. (2018).
\newblock Double/debiased machine learning for treatment and structural parameters.

\bibitem[Consortium, 2015]{10002015global}
Consortium, . G.~P. (2015).
\newblock A global reference for human genetic variation.
\newblock {\em Nature}, 526(7571):68.

\bibitem[Graham et~al., 2021]{graham2021power}
Graham, S.~E., Clarke, S.~L., Wu, K.-H.~H., Kanoni, S., Zajac, G.~J., Ramdas, S., Surakka, I., Ntalla, I., Vedantam, S., Winkler, T.~W., et~al. (2021).
\newblock The power of genetic diversity in genome-wide association studies of lipids.
\newblock {\em Nature}, 600(7890):675--679.

\bibitem[Hao et~al., 2025]{hao2025transfer}
Hao, C., Lei, H., Zhou, X., and Xue, F. (2025).
\newblock Transfer learning for mendelian randomization with pleiotropic and correlated high-dimensional exposures and its application to trans-ethnic populations.

\bibitem[Hayashi, 2011]{hayashi2011econometrics}
Hayashi, F. (2011).
\newblock {\em Econometrics}.
\newblock Princeton University Press.

\bibitem[Hou et~al., 2025]{hou2025temr}
Hou, L., Wu, S., Yuan, Z., Xue, F., and Li, H. (2025).
\newblock Temr: Trans-ethnic mendelian randomization method using large-scale gwas summary datasets.
\newblock {\em The American Journal of Human Genetics}, 112(1):28--43.

\bibitem[Huang et~al., 2026]{huang2026xmr}
Huang, X., Chao, Z., Wang, Z., Hu, X., and Yang, C. (2026).
\newblock Xmr: A cross-population mendelian randomization method for causal inference using genome-wide summary statistics.
\newblock {\em medRxiv}, pages 2026--03.

\bibitem[Lambert et~al., 2021]{lambert2021polygenic}
Lambert, S.~A., Gil, L., Jupp, S., Ritchie, S.~C., Xu, Y., Buniello, A., McMahon, A., Abraham, G., Chapman, M., Parkinson, H., et~al. (2021).
\newblock The polygenic score catalog as an open database for reproducibility and systematic evaluation.
\newblock {\em Nature genetics}, 53(4):420--425.

\bibitem[Li et~al., 2022]{li2022transfer}
Li, S., Cai, T.~T., and Li, H. (2022).
\newblock Transfer learning for high-dimensional linear regression: Prediction, estimation and minimax optimality.
\newblock {\em Journal of the Royal Statistical Society Series B: Statistical Methodology}, 84(1):149--173.

\bibitem[Li et~al., 2023]{li2023estimation}
Li, S., Zhang, L., Cai, T.~T., and Li, H. (2023).
\newblock Estimation and inference for high-dimensional generalized linear models with knowledge transfer.
\newblock {\em Journal of the American Statistical Association}, pages 1--12.

\bibitem[Mikusheva and Sun, 2024]{mikusheva2024weak}
Mikusheva, A. and Sun, L. (2024).
\newblock Weak identification with many instruments.
\newblock {\em The Econometrics Journal}, page utae007.

\bibitem[Mokdad et~al., 2003]{mokdad2003prevalence}
Mokdad, A.~H., Ford, E.~S., Bowman, B.~A., Dietz, W.~H., Vinicor, F., Bales, V.~S., and Marks, J.~S. (2003).
\newblock Prevalence of obesity, diabetes, and obesity-related health risk factors, 2001.
\newblock {\em Journal of the American Medical Association}, 289(1):76--79.

\bibitem[Newey, 1990]{newey1990efficient}
Newey, W.~K. (1990).
\newblock Efficient instrumental variables estimation of nonlinear models.
\newblock {\em Econometrica: Journal of the Econometric Society}, pages 809--837.

\bibitem[Newey, 1993]{newey1993eff}
Newey, W.~K. (1993).
\newblock {\em Efficient estimation of models with conditional moment restrictions}.
\newblock Elsevier.

\bibitem[Newey and Windmeijer, 2009]{newey2009generalized}
Newey, W.~K. and Windmeijer, F. (2009).
\newblock Generalized method of moments with many weak moment conditions.
\newblock {\em Econometrica}, 77(3):687--719.

\bibitem[Ruan et~al., 2022]{ruan2022improving}
Ruan, Y., Lin, Y.-F., Feng, Y.-C.~A., Chen, C.-Y., Lam, M., Guo, Z., He, L., Sawa, A., Martin, A.~R., et~al. (2022).
\newblock Improving polygenic prediction in ancestrally diverse populations.
\newblock {\em Nature Genetics}, 54(5):573--580.

\bibitem[Stock et~al., 2002]{stock2002survey}
Stock, J.~H., Wright, J.~H., and Yogo, M. (2002).
\newblock A survey of weak instruments and weak identification in generalized method of moments.
\newblock {\em Journal of Business \& Economic Statistics}, 20(4):518--529.

\bibitem[Su et~al., 2011]{su2011hapgen2}
Su, Z., Marchini, J., and Donnelly, P. (2011).
\newblock Hapgen2: simulation of multiple disease snps.
\newblock {\em Bioinformatics}, 27(16):2304--2305.

\bibitem[Tian and Feng, 2022]{tian2022transfer}
Tian, Y. and Feng, Y. (2022).
\newblock Transfer learning under high-dimensional generalized linear models.
\newblock {\em Journal of the American Statistical Association}, pages 1--14.

\bibitem[Wang et~al., 2025]{wang2025debiased}
Wang, R., Zhang, H., and Lin, X. (2025).
\newblock Debiased estimating equation method for robust and efficient mendelian randomization using a large number of correlated weak and invalid instruments.

\bibitem[Yang and Zhu, 2025]{yang2025improving}
Yang, Y. and Zhu, X. (2025).
\newblock Improving causal effect estimation in multi-ancestry multivariable mendelian randomization with transfer learning.
\newblock {\em bioRxiv}, pages 2025--07.

\bibitem[Zhang et~al., 2023]{zhang2023new}
Zhang, H., Zhan, J., Jin, J., Zhang, J., Lu, W., Zhao, R., Ahearn, T.~U., Yu, Z., O’Connell, J., Jiang, Y., et~al. (2023).
\newblock A new method for multiancestry polygenic prediction improves performance across diverse populations.
\newblock {\em Nature Genetics}, 55(10):1757--1768.

\bibitem[Zheng et~al., 2022]{zheng2022multi}
Zheng, J., Zhang, Y., Zhao, H., Liu, Y., Baird, D., Karim, M.~A., Ghoussaini, M., Schwartzentruber, J., Dunham, I., Elsworth, B., et~al. (2022).
\newblock Multi-ancestry mendelian randomization of omics traits revealing drug targets of covid-19 severity.
\newblock {\em EBioMedicine}, 81.

\end{thebibliography}
}

\end{document}